\documentclass[openacc]{rsproca_new}

\usepackage{bm}

\def \bA {{\bf A}}
\def \bB {{\bf B}}

\def \bE {{\bf E}}

\def \bj {{\bf j}}

\def \bN {{\bf N}}
\def \bn {{\bf n}}

\def \bu {{\bf u}}
\def \bv {{\bf v}}
\def \bw {{\bf w}}
\def \del {{\bm \nabla}}
\def \half {\textstyle{\frac{1}{2}}}
\def \div {{\bm \del} \cdot}
\def \curl {{\bm \del} \times}
\def \p {\partial}

\def \. {\cdot}
\def \x {\times}
\def \be {\begin{equation}}
\def \ee {\end{equation}}

\begin{document}

\title{Three-dimensional magnetic reconnection in astrophysical plasmas}

\author{
Ting Li$^{1,2}$, Eric Priest$^{3}$ and Ruilong Guo$^{4}$}

\address{$^{1}$CAS Key Laboratory of Solar Activity, National
Astronomical Observatories, Chinese Academy of Sciences, Beijing
100101, China\\
$^{2}$School of
Astronomy and Space Science, University of Chinese Academy of
Sciences, Beijing 100049, China\\
$^{3}$Mathematics Institute, St Andrews University, ST ANDREWS KY16 8QR, UK\\
$^{4}$Laboratory for Planetary and Atmospheric Physics, STAR
institute, Universit\'{e} de Li\`{e}ge, Li\`{e}ge, Belgium }

\subject{Astrophysics, Solar Physics, Plasma Physics}


\corres{Ting Li\\
\email{liting@nao.cas.cn}}

\begin{abstract}
Magnetic reconnection is a fundamental process in a laboratory,
magnetospheric, solar and astrophysical plasma, whereby magnetic
energy is converted into heat, bulk kinetic energy and fast particle
energy. Its nature in two dimensions is much better understood than
in three dimensions (3D), where its character is completely
different and has many diverse aspects that are currently being
explored. Here we focus on the magnetohydrodynamics of 3D
reconnection in the plasma environment of the solar system,
especially solar flares.
 The theory of reconnection at null points, separators
and quasi-separators is  described, together with accounts of
numerical simulations and observations of these three types of
reconnection. The  distinction  between  separator  and
quasi-separator reconnection is a theoretical one that is
unimportant for the  observations of energy release. A new paradigm
for solar flares, in which 3D reconnection plays a central role, is
proposed.
\end{abstract}


\begin{fmtext}
\section{Introduction}
\label{sec1}
Magnetic reconnection is the basic paradigm in astro-physical, space and laboratory plasmas for converting magnetic energy into other forms, namely, heat,
bulk kinetic energy and fast particle energy.
It enables magnetic field lines globally to re-structure by locally changing their connections with one another.
\end{fmtext}
\maketitle In this review we focus on the magnetohydrodynamics (or
MHD) of the process and its application in the solar atmosphere,
especially in coronal heating and solar flares
\cite{pontin12b,longcope15,janvier17}, and we also describe a little
of its action in the Earth's magnetosphere. Determining the
mechanism for heating the Sun's outer atmosphere or {\it corona} to
a million degrees K compared with its surface temperature of 6000 K
represents a major challenge in astrophysics, and solar flares are
the largest and most  complex release of magnetic energy in the
solar system.

 The history of the
study of magnetic reconnection can be traced back to the 1950s.
Dungey \cite{1953MNRAS.113..679D} was the first to suggest that
magnetic field lines  can be disconnected and rejoined in a location where
a strong Ohmic current exists. Several years later, the Sweet-Parker
model\cite{1957JGR....62..509P,1958IAUS....6..123S} was proposed, in
which the magnetic field is dissipated at a large-scale current
sheet surrounding an X-type neutral point where the magnetic field vanishes. However, the efficiency
of reconnection is limited by the weak diffusion of magnetic field
at such a sheet and so the reconnection rate is much too slow for
solar flares.

Afterwards, Petschek\cite{petschek64} realized that the current
sheet can be very much smaller and that slow-mode magnetic shock
waves naturally propagate from its ends  and stand in the plasma
flow. These shocks help to convert magnetic energy into heat and
kinetic energy. The resulting reconnection rate is much higher in
the Petschek model than the Sweet-Parker model, and is indeed rapid
enough for flare energy release. Petschek's model is
``Almost-Uniform" in the sense that the inflow magnetic field is
weakly curved, but, since then, it
 has been generalised to give other fast,  Almost-Uniform,
reconnection regimes, which depend on the boundary conditions and
initial state \cite{priest86a} and  have been well established by
numerical simulations (Sec.2\ref {sec2.1}). Indeed, many other
details of two-dimensional (2D) reconnection in a range of plasma
environments have been studied \cite{priest00,
Ni2020RSPSA.47690867N}.

It is clear that magnetic reconnection is intrinsically a
three-dimensional (3D) process, and so, for the last twenty years,
one of the main foci of researchers has been on the structure and
dynamics of 3D reconnection. Although quasi-2D models have been
highly successful at reproducing many features of reconnection in
the Earth's magnetosphere and in solar flares, it transpires that
fully 3D reconnection operates in other ways that are rich and
varied \cite{Schindler1988,Hesse1988,Priest2003}.

In 3D, there are three types of location where magnetic reconnection can take place since they are natural locations where large currents tend to form, provided the
right flows are present:
\vspace{0.2cm}

 (i) {\it 3D\ null\ points}, where $\bB=0$\cite{Lau1990ApJ...350..672L};

(ii) {\it separator\ field\ lines}, which are the intersections of two separatrix surfaces, across which the magnetic field
connectivity changes in a discontinuous
way\cite{sweet58b
}.

(iii) {\it quasi-separators} \cite{priest1995,Demoulin1996} or {\it hyperbolic flux tubes} \cite{Titov2002JGRA..107.1164T}, which are the intersections of
two quasi-separatrix layers (QSLs) at which magnetic connectivity changes are continuous but rapidly varying.
\vspace{0.2cm}

A key new feature of reconnection in 3D, both at nulls, separators
and quasi-separators, is the presence of {\it magnetic\ flipping} or
counter-rotation, first suggested by Priest \& Forbes
\cite{Priest1992} and observed in solar observations by Mandrini et
al.\cite{Mandrini1991}. Another related feature is that  the
magnetic field lines continually change their connections throughout
the diffusion region, instead of the classical cut-and-paste
reconnection  at a single point that occurs in 2D \cite{Priest2003}.


Solar flares are, like geomagnetic substorms and dayside
reconnection, one of the
 most direct consequences of magnetic
reconnection in our solar system. They emit radiation over the whole
range of the electromagnetic spectrum, with the largest radiative
increase in the extreme ultraviolet (EUV) and soft X-rays. Magnetic
reconnection is now recognized as the process that releases free
magnetic energy stored in the sheared or twisted magnetic fields of
active regions (ARs) during  solar flares\cite{
Shibata2011}. There has been much observational evidence to support this paradigm, including:
\vspace{0.2cm}

(a) reconnection inflows and outflows \cite{Yokoyama2001ApJ,Savage2010ApJ...722..329S,
Sun2015NatCo...6.7598S},

(b) supra-arcade downflows
\cite{McKenzie1999ApJ...519L..93M,Lillp2016ApJ},

(c) cusp-shaped flare loops \cite{Masuda
},

(d) X-ray sources in the current sheet \cite{LinJ2005ApJ...622.1251L,
Xue2018ApJ...858L...4X
}, at loop tops and at the footpoints of flaring loops \cite{Masuda,
Su2013NatPh...9..489S}. \vspace{0.2cm}

In order to explain the different observational signatures of solar
flares, the standard 2D CSHKP model was
developed\cite{
Kopp1976SoPh...50...85K}, in which a large-scale magnetic flux rope
starts to move upward due to a loss of equilibrium \cite{priest90a}
or eruptive instability \cite{Kliem2006PhRvL..96y5002K} and
stretches the overlying magnetic
field. A vertical current sheet is formed  under the rising flux rope 
and magnetic reconnection is driven at it, generating energetic
particles and thermal energy, which propagate downwards along the
reconnected field lines, and then impact  the lower and denser
layers of the solar atmosphere. This produces flare loops and
ribbons in X-rays, EUV, ultraviolet (UV) and chromospheric
wavelengths such as H$\alpha$.

2D (and 2.5D) flare models have been remarkably effective in
explaining many basic aspects of solar flares (Sec.2\ref{sec2.4}).
However, in recent years,  high-resolution imaging and spectroscopic
observations of the Sun (e.g., SDO and IRIS), have revealed more
complex details
 of solar flares that lie outside a 2D or 2.5D picture.
 Many of these new features are being
explained by 3D modelling, such as: the formation of flare ribbons
beginning with small kernels; the creation of twist in an erupting
flux rope; the spread of a flare and the sympathetic triggering of a
sequence of eruptions; the hook-shaped ends of flare ribbons; and
the observed motions of structures along arcades and ribbons. Thus,
flare models involving 3D magnetic reconnection are being developed
and a new flare paradigm has been emerging.

3D reconnection is also important in reconfiguring the Earth's
magnetosphere. Some particle-in-cell simulations and \emph{in situ}
observations from multiple spacecrafts show a role for 3D magnetic
reconnection (Sec.3b(\ref{sec3.2.4})). For example, 3D null points
can sometimes be related to the formation of flux ropes, as well as
energy dissipation and particle acceleration
(Sec.3a(\ref{sec3.1.2})). However, in applications to both the solar
atmosphere and the Earth's magnetosphere, 2D simulations and theory
remain of value and complement 3D modelling, especially concerning
time-dependent and kinetic aspects.

In the following, we review the theory of 3D MHD reconnection
(Sec.\ref{sec2}), as well as 3D computational models and
observations of energy release in solar flares and the Earth's
magnetosphere (Sec.\ref{sec3}), and we summarise the new solar flare paradigm
(Sec.\ref{sec4}).

\section{Theory of Magnetic Reconnection}
\label{sec2}
\subsection{Introduction}
\label{sec2.1} Magnetic reconnection is a fundamental process in
plasmas throughout the Universe, by which magnetic field lines
change their connections and magnetic energy is converted into heat,
kinetic energy and fast particle energy. It lies at the core of
solar flares  and geomagnetic substorms. Here we focus on its
occurrence in a plasma for which MHD is valid and for which the
global {\it magnetic\ Reynolds\ number} is much larger than unity,
i.e.,
\begin{equation}
R_{me}=\frac{L_ev_{e}}{\eta}   \gg 1,
\label{Rm}
\end{equation}
where $L_e$ is the global (i.e., external) length-scale, $v_e$ is
the global plasma speed, $B_e$ the global magnetic field, and $\eta$
the magnetic diffusivity. $R_m$ is sometimes based on the global
{\it Alfv\'{e}n\ speed} ($v_{Ae}=B_e/\sqrt{\mu \rho}$) rather than
the flow speed $v_{e}$, when it may also be called the {\it
Lundquist\ number}. One of the MHD equations is the {\it induction\
equation}
\begin{equation}
\frac{\partial {\bf B}}{\partial t}={{\bf \nabla} \times}({\bf v}\times{\bf B})+\eta \nabla^2 {\bf B},
\label{indeq}
\end{equation}
and another is the {\it equation\ of\ motion}
\begin{equation}
\rho \frac{d \bf v}{dt}=-{\bf\nabla}p + \bf j \times B,
\label{motion}
\end{equation}
where $\div \bB =0$, $\rho$ is the plasma density, ${\bf j} = {\bf
\nabla} \times {\bf B}/\mu$ the electric current density, $p =
{\tilde{R}} \rho T$ the plasma pressure, $\tilde{R}$ the gas
constant, and the temperature ($T$) is determined by an energy equation.

Condition (\ref{Rm}) ensures that the second term on the right-hand
side of (\ref{indeq}), namely, magnetic diffusion, is negligible in
most of the volume,  so that the magnetic field is ``frozen into the
plasma" and none of its energy converts into heat by ohmic
dissipation.
 However, in extremely narrow
regions, often sheets, where the electric current is so strong and
the length-scale so small that $R_m \approx 1$, the magnetic
diffusion term is important and the magnetic field can slip through
the plasma. In this case, the magnetic field lines change their
connections, magnetic reconnection occurs, and magnetic energy is
converted into other forms, often accelerating hot fast jets of
plasma away from the reconnection site.

The theory in 2D \cite{priest14a}  shows how fast reconnection at
typically $0.001-0.1 v_{Ae}$ occurs in three different situations:
\vspace{0.2cm}

(i) by {\it Almost-Uniform reconnection} \cite{priest86a} when the magnetic diffusivity is enhanced in the central current
sheet \cite{baty14}; this is a generalisation of Petschek's mechanism \cite{petschek64}, in
which slow-mode shock waves stand in the flow and radiate from a tiny central current sheet;

(ii) by {\it impulsive\ bursty\ reconnection} when the current sheet is long enough that it goes unstable to
secondary tearing \cite{priest86b,forbes87,loureiro07,bhattacharjee09};

(iii) or by {\it Hall reconnection} when the medium is collisionless
and the Hall effect is important, with a resistive diffusion region
replaced by an ion diffusion region surrounding an electron
diffusion region \cite{shay98b}.
\vspace{0.2cm}

In 3D we are in a process of discovery about a new territory, since
it transpires that 3D reconnection is very different from 2D
reconnection in many ways (Sec.2\ref{sec2.4}). But, before
describing them we need to give some background about geometry,
topology, flux and field-line conservation (Sec.2\ref{sec2.2}), and
conditions for reconnection (Sec.2\ref{sec2.3}). Then we shall be
ready to describe briefly the different regimes of reconnection,
namely, null-point reconnection (Sec.2\ref{sec2.5}), separator
reconnection (Sec.2\ref{sec2.6}), and quasi-separator reconnection
(Sec.2\ref{sec2.7}).

\subsection{Background Concepts: Geometry and Topology}
\label{sec2.2} The structures of the magnetic field around null
points in 3D include isolated field lines called {\it spines}  and a
surface of field lines called a {\it fan}, which originate or end at
the null, nomenclature that was coined by Priest and Titov
\cite{Priest1996Titov}. For a {\it positive null} the field lines
enter the null along the spine and leave it in the fan, while for a
{\it negative null} they enter in the fan and leave along the spine
(see also Lau and Finn\cite{Lau1990ApJ...350..672L}, who earlier
used a different notation, namely, B-type for positive nulls, A-type
for negative nulls, and As-type or Bs-type for spiral nulls). The
simplest example of a positive null has magnetic field components
\begin{equation}
    (B_x,B_y,B_z) = (x,y,-2z)
    \label{eq_null}
\end{equation}
that satisfy $\div \bB=0$, as shown in Fig.{\ref{fig_null}a. This is
a {\it proper\ radial\ null}, for which the spine is perpendicular to the
fan, and the field lines in the fan are straight.

\begin{figure}[!h]
\centering
\includegraphics[height=1.5in]{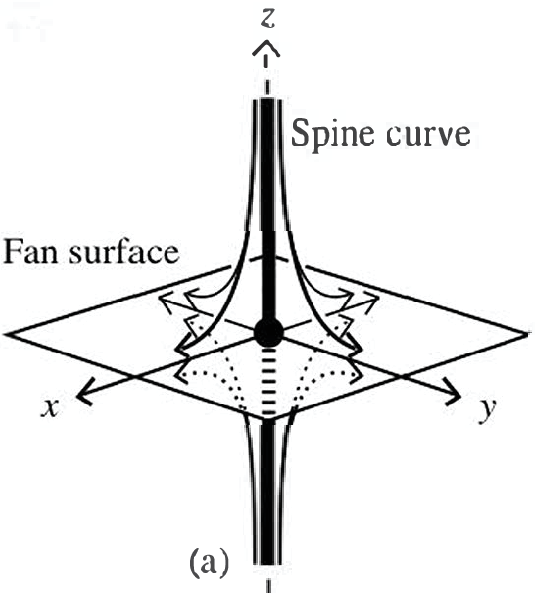}
\includegraphics[height=1.5in]{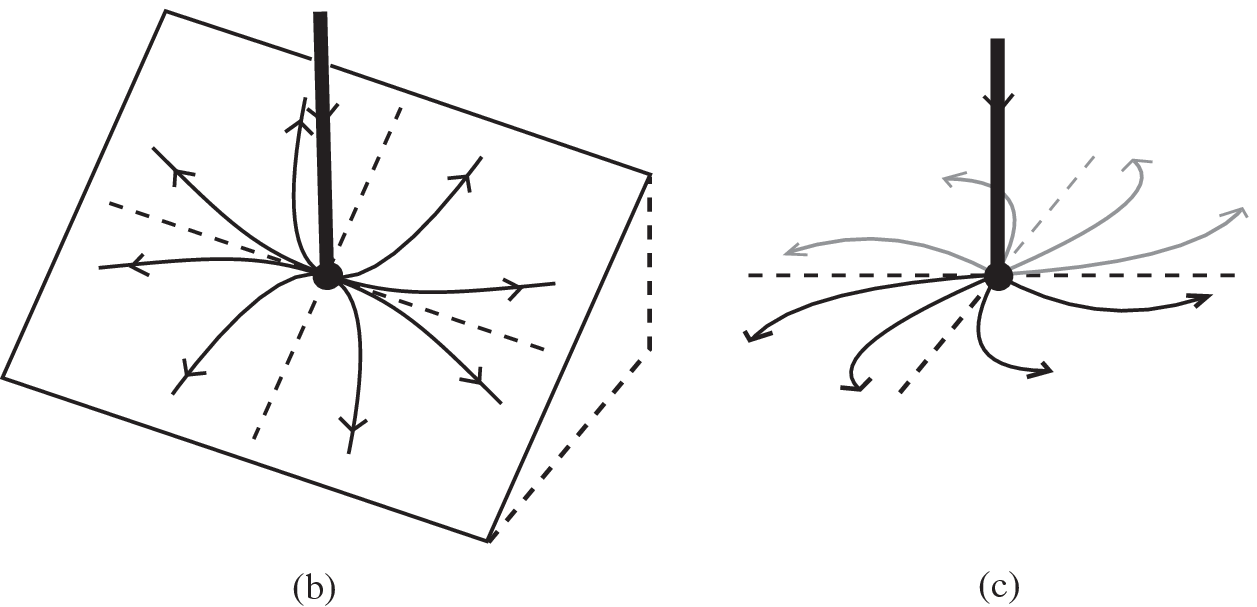}
 \caption{(a) The
nature of magnetic field lines near a null point in 3D with a spine
and a fan, for (a) a proper radial null, (b) an oblique null, and
(c) a spiral null (from Priest\cite{priest14a} with permission).}
\label{fig_null}
\end{figure}
The most general form of a {\it linear\ null}, for which the field components increase linearly away from the null, may be written \cite{Parnell1996}
\be
\left( \begin{array}{c}
B_{x} \\ B_{y} \\ B_{z} \end{array}\right) = \left(
\begin{array}{ccc}
1 & \frac{1 }{ 2}(b-j_{\parallel}) & 0 \\
\frac{1 }{ 2} (b+j_{\parallel}) & a & 0 \\
0 & j_{\perp} & -a-1 \end{array} \right) \left( \begin{array}{c} x
\\ y \\ z \end{array}\right). \label{eq_gennull} \ee This includes
both {\it oblique nulls}, in which the spine and fan are no longer
perpendicular, and {\it spiral nulls}, for which the field lines in
the fan spiral inwards or outwards.  Null points are common in the
corona \cite{
edwards15a}, where they often appear at the
summit of a {\it separatrix dome} that lies above a region of
photospheric parasitic polarity surrounded by a region of the
opposite polarity (Fig.\ref{fig_domes}a).

\begin{figure}[!h]
\centering\includegraphics[height=1.5in]{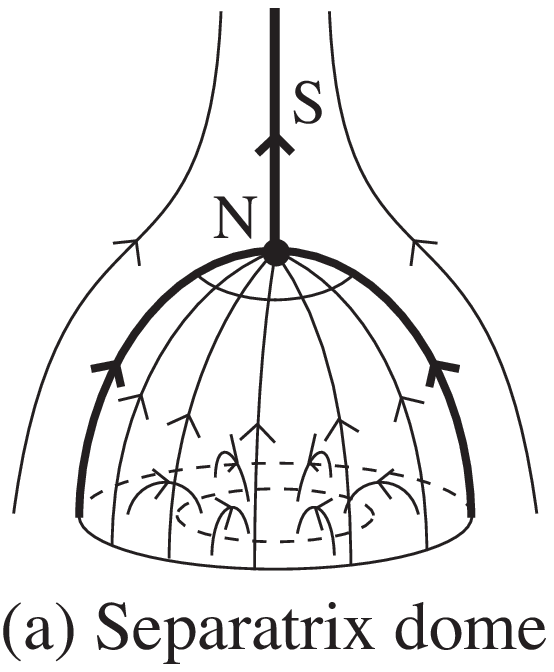}
\centering\includegraphics[height=1.5in]{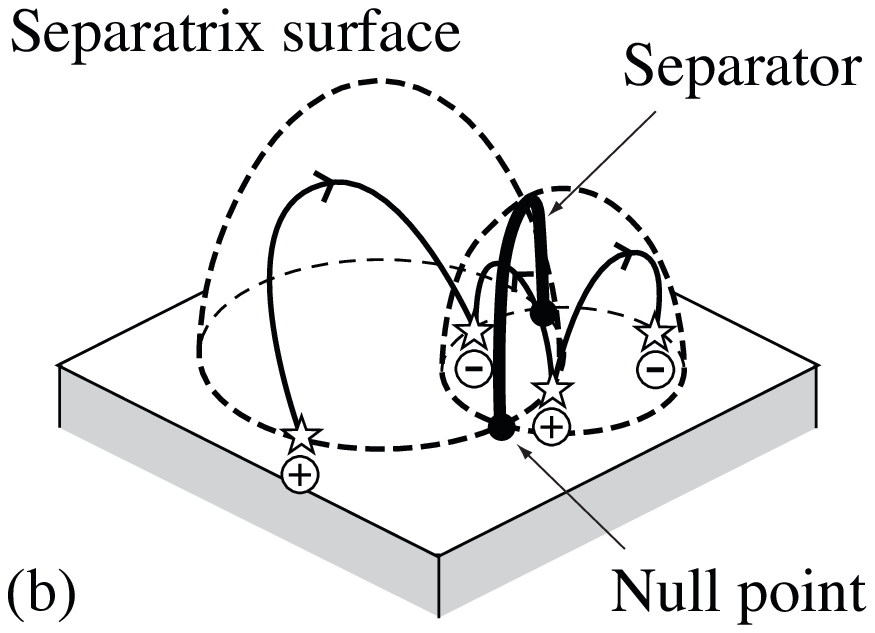} \caption{The
topology of (a) a separatrix dome with a coronal null point N lying
above a region of parasitic polarity and (b) two separatrix domes
that intersect in a separator and lie above two sources of positive
and two of negative polarity (from Priest\cite{priest14a} with
permission).} \label{fig_domes}
\end{figure}

The {\it skeleton} of a magnetic field consists of the {\it
separatrix surfaces} (or {\it separatrices}), across which the
mapping of field lines from their footpoints in the solar surface is
discontinuous. Such separatrices originate either in so-called {\it
bald patches}, where the magnetic field touches the boundary
\cite{titov93a}, or, more usually, at the fans of null points.
Separatrices intersect in special field lines called {\it
separators} that usually link one null point with another. Such
separators were first considered by Sweet\cite{sweet58b} and later
analyzed by many others
\cite{Lau1990ApJ...350..672L,Priest1996Titov,
Longcope1996,
parnell10a}. An example is shown in
Fig.\ref{fig_domes}b, where four flux sources (two positive and two
negative) in the solar surface produce two separatrix domes that
originate as the fans of two null points, also lying in the solar
surface, and the domes intersect in the separator. The skeleton can
therefore be mapped out by determining the null points, the fan
surfaces and the separators, an efficient method for which has been
developed by Haynes et al.\cite{haynes07}.

The {\it quasi-skeleton}, in contrast, consists of {\it
quasi-separatrix layers (QSLs)} \cite{priest95a}, which are surfaces
across which the gradient of the mapping of field lines from one
footpoint to another (Eq.\ref{eq_map}) is not singular but is very
large. If a weak uniform field is added to a field with
separatrices, the null points become locations where the field is
weak but non-vanishing and the remnants of the separatrices become
QSLs. The QSLs intersect in {\it quasi-separators} and are best
diagnosed by mapping out the regions where the so-called {\it
squashing\ degree} $Q$ is large, defined as follows
\cite{Titov2002JGRA..107.1164T}. First, split the surface of the
volume under consideration into parts $S_0$ and $S_1$ where the
field lines enter the volume at points $(x_0,y_0)$ and leave it at
$(x_1,y_1)$ in cartesian coordinates. Then, the displacement
gradient tensor is formed from the gradients of the mapping
functions $x_1(x_0,y_0)$, $y_1(x_0,y_0)$  as
\begin{equation}
{\bf \cal{F}}=
\left( \begin{array}{ll}
 \partial x_1/\partial x_0 & \partial x_1 / \partial y_0\\
\partial y_1 / \partial x_0 &  \partial y_1 / \partial y_0
\end{array} \right),
\label{eq_map}
\end{equation}
while $Q$ is defined as
\begin{equation}
Q=-\frac{B_{z+}}{B^{*}_{z-}}
\left[ \left(\frac{\partial x_1 }{ \partial x_0} \right)^{2} +
\left( \frac{\partial x_1 }{ \partial y_0} \right)^{2} + \left(
\frac{\partial y_1 }{ \partial x_0} \right)^{2} + \left( \frac{\partial y_1
}{ \partial y_0} \right)^{2} \right]
\label{eq_Q}
\end{equation}

where $B_{z+}(>0)$ and
$B^{*}_{z-}(<0)$ are the normal components of the field at the two ends of a field line.
}
The region around a quasi-separator where Q is highest is called a  {\it
hyperbolic flux tube (HFT)} \cite{Titov2002JGRA..107.1164T}. It is
bounded by a flux surface $Q=\mbox{const}\gg 1$ with a shape that
continuously changes along the HFT from a narrow flattened tube to a
cross and then to a another narrow flattened tube at the other end
and perpendicular to the first one, as follows
\cite{Titov2002JGRA..107.1164T}:
 $$
\begin{picture}(125,8)(0,0)
  \multiput(0,0)(-0.3,0){6}{\line(1,1){10}}
      \put(14,2){$\rightarrow$}
  \multiput(28.15,2)(-0.28,0){5}{$\times$}
  \multiput(28,0)(-0.3,0){6}{\line(1,1){10}}
      \put(42,2){$\rightarrow$}
  \multiput(56,0)(-0.3,0){6}{\line(1,1){10}}
  \multiput(66,0)(-0.3,0){6}{\line(-1,1){10}}
      \put(71,2){$\rightarrow$}
  \multiput(95,0)(-0.3,0){6}{\line(-1,1){10}}
  \multiput(85.15,2)(-0.28,0){5}{$\times$}
      \put(100,2){$\rightarrow$}
  \multiput(124,0)(-0.3,0){6}{\line(-1,1){10}}
\end{picture}.
   $$

\begin{figure}[!h]
\centering\includegraphics[width=6.5 cm]{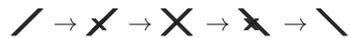}
\centering\includegraphics[width=6.5 cm]{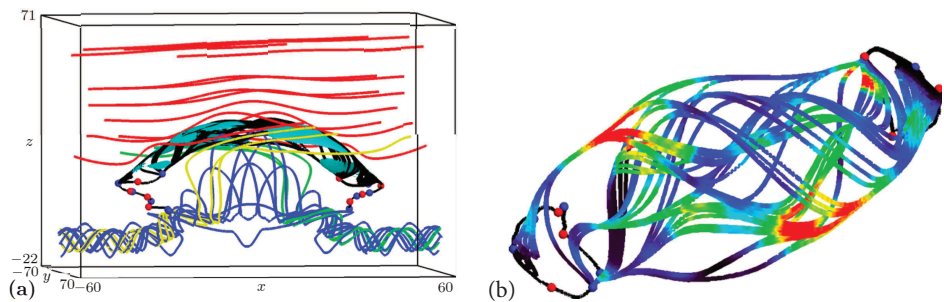}
\caption{A numerical simulation for the emergence of a flux tube
through the solar surface, showing (a) magnetic field lines viewed
from the side, with  separators (thick black lines) linking nulls,
and (b) the separators and nulls viewed from above  (from Parnell et
al.\cite{parnell10b} with permission).} \label{fig_separators}
\end{figure}
Note that maps of Q will reveal both the skeleton and
quasi-skeleton, but do not distinguish between them, and so, in
order to isolate the quasi-skeleton, one needs first to determine
the skeleton separately. Indeed, there are often more null points
and separators in configurations than are intuitively expected. For
example, a computational experiment on new magnetic flux emerging
through the photosphere possessed 18 nulls  and 229 separators
\cite{parnell10b} (Fig.\ref{fig_separators}).

An important topological quantity in 3D reconnection studies is the
{\it magnetic\ helicity}, part of which (the {\it self-helicity})
represents the twisting and kinking of flux tubes, while the
remainder (the {\it mutual\ helicity}) measures the linkage between
flux tubes. Its significance lies in the fact that it is conserved
in an ideal medium and decays extremely slowly in a resistive one,
so that only a very small change takes place in a reconnection event
\cite{woltjer58}. Thus, its approximate conservation is a constraint
on the evolution of coronal magnetic fields \cite{heyvaerts84}.

For a magnetic field $\bB=\curl \bA$,  a gauge-invariant expression for the {\it relative\ magnetic\ helicity} \cite{berger84b} inside a volume V bounded by a surface S is given by
\be
 H =\int_{V_{\infty}} ({\bf A \cdot B - A}_{0} \cdot {\bf B}_{0})\; dV, \label{eq_H_m}
\ee
where ${\bf B}_{0} = {\bf {\bm \nabla} \times A}_{0}$ is current-free inside $V$, with  $\bB_0=\bB$ outside $V$ and  ${\bf A \times n = A}_{0} \times {\bf n}$ on $S$.  Its rate of change due to motions on a boundary (S) is
\be
\frac{dH}{dt} = -2 \int_{V} ({\bf j \cdot B} / \sigma) \; dV + 2 \int_{S} [({\bf
B \cdot A}_{p})  ({\bf v \cdot n) - (v \cdot A}_{p}) ({\bf B \cdot n})] \; dS,
\label{eq_H_mchange}
\ee
when the gauge ($\bA_p$) is chosen such that $\div \bA_p=0$ and $\bA_p\cdot \bn=0$ on S. The first term represents helicity dissipation, which is very small, while the others give the transport of helicity
through the surface, the first by flux emergence and the second due to footpoint motion.

Later, by extending earlier ideas of Berger\cite{berger88}, Yeates
\& Hornig\cite{yeates13} proposed a new topological flux function
(the {\it field\ line\ helicity}) for magnetic field lines that
connect two boundaries: \be {\cal A}=\int {\bA\cdot} d{\bf s},
\label{eq_fl_hel} \ee with the integral being taken from one end of
a field line to the other end. It is an ideal invariant and measures
the average poloidal flux around a field line. Its integral over one
of the boundaries gives the relative magnetic helicity of the
volume. It was  applied to the evolution of coronal magnetic fields
and shown to be concentrated in flux ropes \cite{yeates16}.

\subsection{Conditions for Reconnection}
\label{sec2.3} In an ideal plasma, when $R_m \gg 1$, Ohm's law and
its curl, namely, the induction equation, become \be \bE+\bv \x \bB=
\bf 0 \label{eq_idealohm} \ee and \be \frac{\p \bB}{\p t}=\curl (\bv
\x \bB), \label{eq_idealindeq} \ee which implies that both magnetic
flux and magnetic field lines are conserved, i.e., plasma elements
that form a flux tube or are joined by a field line will continue to
do so.

However, in a non-ideal plasma, although flux conservation implies
field-line conservation, the opposite is not true. Suppose in this
case Ohm's Law takes the form \be \bE+\bv \x \bB= \bN,
\label{eq_Ohmgen} \ee with a general nonideal term $\bN$. A
variation of a magnetic field then preserves magnetic flux if a {\it
magnetic flux velocity} ($\bw$) exists that satisfies \be \frac{\p
\bB}{\p t}=\curl (\bw \x \bB), \label{eq_idealindeq_w} \ee for which
Faraday's law ($\p \bB/\p t=-\curl \bE$) implies that $\bN$ must
have the form \be \bN=\bu \x \bB+\del \Phi, \label{eq_N} \ee where
$\bu=\bv-\bw$ is the {\it slippage velocity} and $\Phi$ is a
potential. The presence of $\Phi$ can lead to a component
($E_\parallel)$ of $\bE$ along the magnetic field which is
associated with 3D reconnection.

Comparing Eqns. (\ref{eq_Ohmgen}) and (\ref{eq_N}), we see that \be
\bE+\bw \x \bB=\del \Phi, \label{eq_Ewphi} \ee so that, if
$\bu\cdot\bB=0$, the flux velocity may be written \be
\bw=\bv+\frac{(\bN-\del \Phi)\x\bB}{B^2}, \label{eq_w} \ee which
tends to be singular at null points.

The condition for reconnection depends on the nature of the
non-ideal term $\bN$ in Eq.(\ref{eq_Ohmgen}), as follows
\cite{hornig96}: \vspace{0.2cm}

(a) if $\bN = \bu \x \bB + \del \Phi$,
 ${\bu}$ is smooth, and there is no reconnection, but the magnetic field slips through the
plasma;

(b) if ${\bf N} = {\bf u} \times {\bf B} + \del \Phi$ and ${\bf u}$
is singular, then reconnection occurs in 2D;

(c) if ${{\bf N}} \neq {\bf u}\times {\bf B} + \del \Phi$, then
reconnection occurs in 2.5D or 3D. \vspace{0.2cm}

In case (a) there is a unique smooth flux velocity $\bw$ which
describes the transport of field lines by a unique smooth flow that
preserves the field line topology. However, in 3D reconnection at an
isolated diffusion region, there is no unique $\bw$, and so two flux
velocities ($\bw_{in}$ and $\bw_{out}$) are needed to describe the
behaviour of field lines, depending on whether the field lines are
regarded as attached to the plasma in the ideal region on the side
of the diffusion region where they enter or where they leave. For
the Parker braiding problem or for quasi-separator reconnection, the
flux velocity is non-unique but smooth, whereas for null-point or
separator reconnection it is non-unique and non-smooth.

\subsection{Reconnection in 3D versus 2D}
\label{sec2.4} The nature of reconnection in 3D is completely
different from 2D, since most of the basic properties of 2D
reconnection do not carry over into 3D \cite{Priest2003}. By ``2D
reconnection" we mean reconnection in a strictly two-dimensional
field ($B_x(x,y), B_y(x,y)$) that varies in two dimensions, whereas
``3D reconnection" refers to reconnection in a fully 3D field
($B_x(x,y,z), B_y(x,y,z), B_z(x,y,z)$). Thus, 2D should not be
confused with 2.5D, which we do not treat here and which refers to a
field of the form ($B_x(x,y), B_y(x,y), B_z(x,y)$) with a guide
field ($B_z(x,y)$). A 2D null point is topologically stable in 2D,
in the sense that, if a purely 2D perturbation is made, the null
point will continue its existence and just move its location
slightly in 2D. Similarly, a 3D null point is also topologically
stable. However, a 2.5D field, which exists in 3D, can be
topologically unstable: for example, when a general 3D perturbation
is made to a 2.5D X-line (consisting of a continuum of 2D X-points
stacked on top of one another), it does not remain as an X-line but
breaks up into a pair of 3D null points. There have been many very
useful theories and simulations in 2D and 2.5D, which have helped to
clarify our understanding of reconnection, but most examples in
nature are three-dimensional and so the 2D and 2.5D understanding is
often likely to be only partial.

In 2D, the properties are: \vspace{0.2cm}

(i) Reconnection takes place only at X-type null points, where the magnetic field vanishes and the nearby field has a hyperbolic structure;

(ii) Magnetic flux moves at the flux velocity ($\bw$), which is singular at the X-point;

(iii) The mapping of field lines near an X-point from one footpoint to another is discontinuous as the footpoint crosses a separatrix;

(iv) During their passage through the diffusion region, field lines preserve their connections, except at the X-point, where they break and change their connections;

(v) When part of a flux tube is passing through a diffusion region, its two wings  outside the diffusion region  move with the plasma ($\bw=\bv$), while the  segment inside the diffusion region slips through
the plasma ($\bw \neq \bv$).
\vspace{0.2cm}

In contrast, the properties of 3D reconnection are
\vspace{0.2cm}

(i) Reconnection occurs at null points, but also at separators and quasi-separators;

(ii)  In general the notion of a flux tube velocity ($\bw$) fails;

(iii)  At null points, separatrices and separators, the mapping of field lines from one boundary to another is discontinuous, but at quasi-nulls, quasi-separatrices and quasi-separators it is continuous;

(iv)  During their passage through a diffusion region,  field lines change their connections  continually;

(v)  When a field line is partly in the diffusion region, with one end moving with the plasma, the other end flips through the plasma  with a velocity that is different from the plasma velocity.
\vspace{0.2cm}

In 3D, reconnection is defined as a change in the magnetic
connectivity of plasma elements and, in terms of the electric field
component ($E_{\parallel}$) parallel to the magnetic field,
 it may be diagnosed by the condition
\be
\int E_{\parallel} \neq 0,
\label{eq_recdef}
\ee
 so that the electric field is non-zero along a magnetic field line that is reconnecting \cite{Schindler1988,Hesse1988}.

 Suppose Ohm's law holds and the plasma flow vanishes on the boundary.  Then the change of magnetic helicity (Eqn.\ref{eq_H_mchange}) becomes
 \be
 \frac{dH}{dt}=-2\int \bE \cdot \bB\ dV,
 \nonumber
 \ee
 but $\bE \cdot \bB=0$ outside the diffusion region $D_R$, and so this reduces to
 \be
 \frac{dH}{dt}=-2\int_{D_R} E_{\parallel}B\ dV.
 \nonumber
 \ee
 Thus, the condition for the magnetic helicity to change in time is identical to the condition that 3D reconnection exists. However, this change in magnetic helicity is extremely tiny compared with the
 total magnetic helicity present, so that, to a high degree of approximation, the total magnetic helicity remains constant.

 The reason that nulls, separators and quasi-separators are associated with reconnection at diffusion regions that surround them is that they are natural locations where strong currents form.
The consequences of reconnection are magnetic flipping and
counter-rotation of field lines caused by a very small change of
magnetic helicity. Other possible physical effects are acceleration
of plasma jets by the resulting strong Lorentz forces, heating of
plasma by Ohmic heating, and, beyond resistive MHD, acceleration of
fast particles by the resulting strong electric fields, turbulence
and shock waves.

\subsection{Null-Point Reconnection: Theory}
\label{sec2.5}
Reconnection can occur in three main ways at a null point, depending on the nature of the flows that are present. The most common form of null-point reconnection is  {\it spine-fan reconnection}  \cite{Priest2009PhPl...16l2101P}, produced
by shearing motions that produce flows across both the spine and fan of the null. On the other hand, twisting motions can produce either {\it torsional spine reconnection} or {\it torsional fan reconnection}.
These modes of reconnection were discovered on the basis of kinematic models and computational experiments.

Steady-state kinematic models for the ideal region may be set up by solving
\begin{equation}
\bE+\bv \x \bB={\bf 0}   \ \ \ {\rm and} \ \ \   \curl \bE = {\bf 0}, \ \ \
\label{eq_kinematic}
\end{equation}
for the velocity $\bv$ and electric field $\bE=\del \Phi$ with a given magnetic field containing a null point \cite{Priest1996Titov}. Corresponding models for an isolated diffusion region were set up
by Hornig \& Priest \cite{hornig03} by solving Ohm's law
\begin{equation}
\bE+\bv \x \bB=\eta\ \curl \bB.
\label{eq_ohm}
\end{equation}
The component of Eqn.(\ref{eq_ohm}) along the magnetic field gives the potential $\Phi$ everywhere as
\begin{equation}
\Phi =\int \frac{\eta\ \bj \. \bB}{B}\ ds +\Phi_{e},
\label{eq_phi}
\end{equation}
namely, an integral along field lines in terms of the imposed value
($\Phi_e$) at one end.  The component of Ohm's law normal to the
magnetic field then determines the flow component $\bv_{\perp}$
normal to the magnetic field as
\begin{equation}
\bv_{\perp}=\frac{(\del \Phi -\eta\ \bj) \x \bB}{B^{2}},
\label{eq_vperp}
\end{equation}
while the rate of change of magnetic flux ($F_m$) (i.e., the reconnection rate) follows from
\begin{equation}
\frac{dF_m}{dt} =\int E_{\parallel} ds.
\label{eq_rateflux}
\end{equation}

\begin{figure}[!h]
\centering (a)\includegraphics[height=.25\textheight]{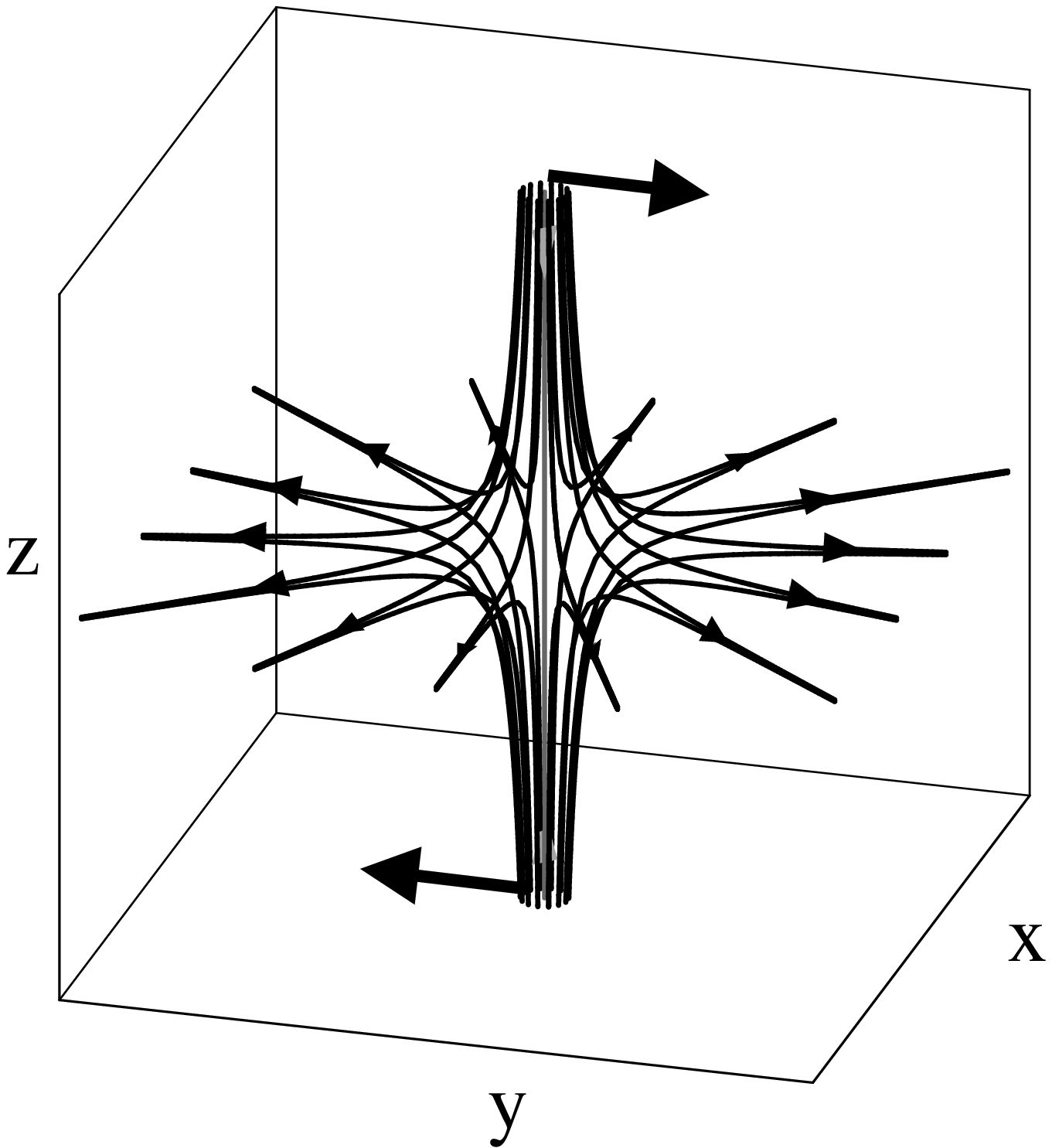} (b)\includegraphics*[height=.25\textheight]{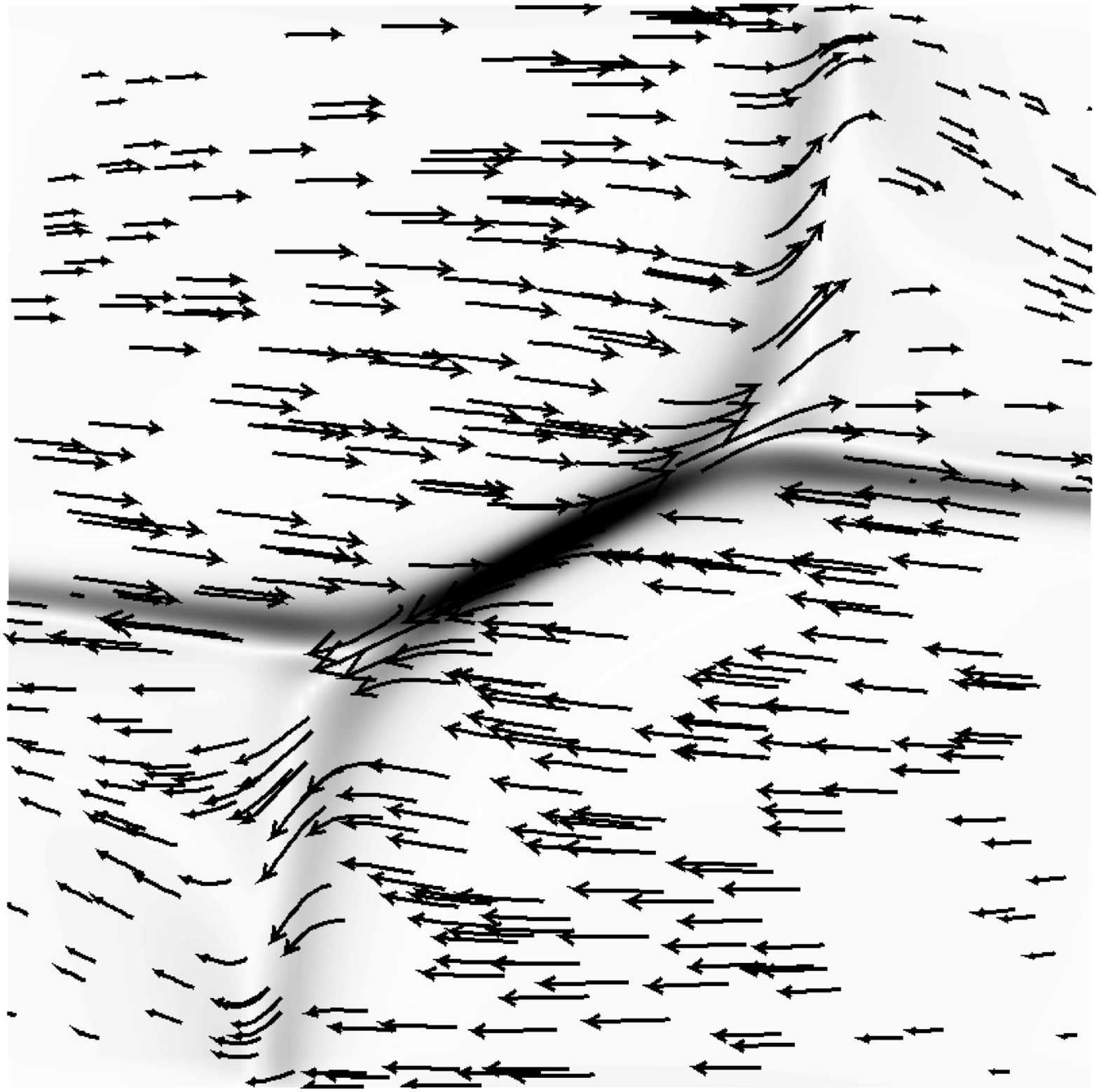}
\caption{(a) An initial shearing in the $y$-direction of a spine aligned with the $z$-axis. (b) The resulting collapse of spine and fan to form {\it spine-fan\ reconnection}, showing the
current-density contours (shaded) and flow velocity arrows in the $x=0$ plane (from Pontin \& Galsgaard\cite{pontin07a} with permission).}
\label{fig_spinefan}
\end{figure}
Pontin et al. \cite{Pontin2005GApFD..99...77P} applied the above approach to a diffusion region in the shape of a disc containing a uniform current along the fan. They found that the plasma flow crosses both
the spine and fan. The field lines flip up and down the spine and around the spine in the plane of the fan. A numerical solution of the full resistive MHD equations by Pontin \& Galsgaard \cite{pontin07a} showed how a
shearing of the spine drives such spine-fan reconnection (Fig. \ref{fig_spinefan}) with a strong fan current.

The kinematic formalism may also be applied to a spiral null with a cylindrical diffusion region. Rotation of the fan plane then tends to drive current along the spine and create twisting motions around the
spine in {\it torsional spine reconnection}.
Inside the diffusion region, rotational slippage allows the field lines to become disconnected and to rotate around the spine. On the other hand, rotation of  the region around the
spine in opposite directions above and below the null drives a strong fan current in  {\it torsional fan reconnection}.  In this case, inside the diffusion region the field lines exhibit
rotational slippage above and below the fan plane.

\subsection{Separator Reconnection: Theory}
\label{sec2.6}
\begin{figure}[!h]
\centering \includegraphics[width=7cm]{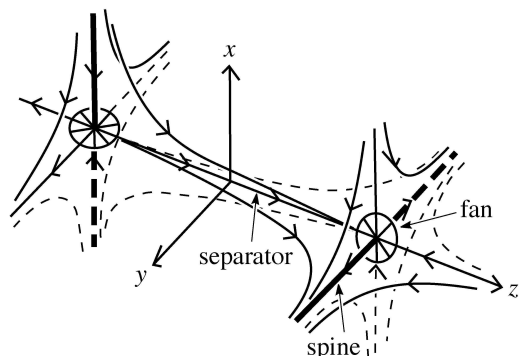}
\caption{An example of a separator that joins two nulls (from Priest\cite{priest14a} with permission).}
\label{fig_separator}
\end{figure}

In complex magnetic fields, null points are common and the particular magnetic field line where the fans of two nulls intersect, called a {\it separator}, is a prime location for the build up of
currents and therefore for reconnection.  For example, the configuration with magnetic field
\be
B_x = x(z-3),\\\ B_y= y(z+3),\ \ \ B_z= 1- z^{2}
\nonumber
\ee
has two nulls. As indicated in Fig. \ref{fig_separator}, one null at $(0,0,-1)$ has its fan orientated in the $yz$-plane, while the other at $(0,0,1)$ has a fan in the $xz$-plane, so that these two fans intersect in the $z$-axis.

In two pioneering papers Sweet \cite{sweet58b}, Lau and Finn \cite{Lau1990ApJ...350..672L} suggested that current sheets can form along separators and lead to reconnection, an idea later developed by others
  \cite{Priest1996Titov,Longcope1996,galsgaard00b}.
A series of computational experiments by Parnell, Longcope and their
colleagues demonstrated the importance of separator reconnection in
coronal heating and in solar flares, as described in
Sec.3\ref{sec3.2}. Furthermore, the relation between separators and
quasi-separators was clarified \cite{restante09}, by showing that
spines and certain portions of fans are good predictors for QSL
footprints and flare ribbons.

For a topologically complex magnetic field, Parnell et al.
\cite{parnell08} discovered the importance of analyzing the skeleton
of separatrix surfaces that spread out from the fans of the null
points. This enabled them to determine how multiple reconnecting
separators can be born  joining the same two null points, and how in
{\it recursive reconnection}  the same flux can be closed and opened
many times. Later, Parnell et al. \cite{parnell10a} were surprised
to find that the magnetic field in a plane perpendicular to a
separator can be either hyperbolic (X-type), as expected, or
elliptic (O-type), and that this  may vary along the separator, with
reconnection occurring at both types of structure.

Longcope \cite{longcope01} also studied how separator current sheets form and dissipate. He demonstrated how the current and energy storage are produced by a change in magnetic flux,
 and
applied the ideas to X-ray bright points \cite{longcope98} and to solar flares \cite{
longcope07a}.

More recently, in order to develop a new model for coronal heating
by flux cancellation, Priest \& Syntelis \cite{priest21a} developed
a method to calculate 2D and axisymmetric 3D separator current
sheets and their reconnecting properties without resorting to
complex variable theory.

\subsection{Quasi-Separator Reconnection: Theory}
\label{sec2.7}
As described in Sec.2\ref{sec2.2},   quasi-separators
or hyperbolic flux tubes are intersections of quasi-separatrices,
which are regions where the mapping gradient (Eq.\ref{eq_map}) or squashing degree
($Q$, Eq.\ref{eq_Q}) is not infinite but is much larger than unity. Since they may often
be regarded as remnants of separators, it is not surprising that
current sheets will tend to form at them and so reconnection is
likely to take place \cite{priest95a}.

\begin{figure}[!h]
\centering \includegraphics[height=1.5in]{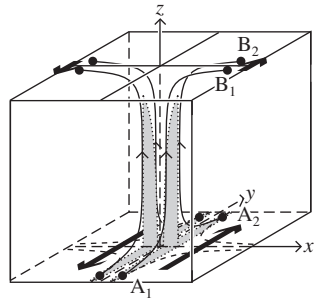}
\caption{An example of a quasi-separator (from Priest\cite{priest14a} with permission).} \label{fig_QSL}
\end{figure}
Consider, for example, a magnetic field with cartesian components
$(x,-y,\epsilon)$, consisting of a uniform $z$-component of
magnitude $\epsilon\ll 1$ superposed on an X-point field in
$xy$-planes (Fig.\ref{fig_QSL}). The field line that maps a point
B$(x_1,y_1,1)$ on the plane $z=1$ to A$(x_0,y_1,0)$ on the plane
$z=0$ is given by \be x_0=x_1 \exp(-1/\epsilon), \ \ \ \ y_0=y_1
\exp(1/\epsilon). \nonumber \ee Thus, the mapping is continuous, but
suppose B moves from B$_1$ to B$_2$ across the $x$-axis with $x_1$
fixed and positive, while $y$ increases from a small negative value
at B$_1$ to a small positive value at B$_2$. Then A will move from a
large negative value at A$_1$ to a large positive value at A$_2$. On
the other hand, if $x_1$ is fixed and negative, A will move in the
opposite direction.  In other words, small motions of the footpoints
on the top boundary across the QSL (the $xz$-plane) produce
extremely rapid flipping of the feet on the bottom boundary. When
magnetic reconnection occurs within QSLs in 3D, field lines exchange
their connectivity with those of their neighbours in the reconnection
layer, and the magnetic field lines  flip (or slip or
slip-run) past each other at super-Alfv\'{e}nic speeds
\cite{Priest1992,priest1995,Aulanier2006SoPh..238..347A}.
Physically, therefore, the behaviour of quasi-separator reconnection
and separator reconnection is very similar on MHD time scales.

Indeed, D{\'e}moulin et al. \cite{Demoulin1996} showed that, if a quasi-separator is present in the corona, the effect of any smooth motion of the photospheric footpoints will be to
create a current sheet there. Also, Titov et al. \cite{titov03a} demonstrated that a stagnation-point flow near a QSL generates strong currents near it, while Aulanier et al. \cite{aulanier05a} and others
confirmed the effect with resistive numerical experiments (see Sec.3\ref{sec3.3}).

\section{Modelling and Observations of 3D Magnetic Reconnection}
\label{sec3}
\subsection{Null-Point Reconnection: Modelling and Observations}
\label{sec3.1} 3D null points as described in Sec.2\ref{sec2.2}
 are preferential
sites for current accumulation and energy dissipation. They have
been observed directly in the Earth's magnetosphere in the
magnetotail\cite{Xiao2006NatPh...2..478X,
Guor2013JGRA..118.6116G} and the polar cusp
region\cite{Dorelli2007JGRA..112.2202D} (Sec.3a\ref{sec3.1.2}).
They are also abundant in the solar atmosphere and a common  feature of solar flares\cite{Longcope1996SoPh..169...91L},
CMEs\cite{Lynch2008ApJ...683.1192L
}, solar jets\cite{Raouafi2016SSRv..201....1R,2021RSPSA.47700217S}
and flux emergence\cite{Torok2009ApJ...704..485T
} (Sec.3a\ref{sec3.1.1}). In particular, they arise commonly when
parasitic polarity surrounded by opposite polarity of greater flux
forms a coronal null  \cite{
Liuw2011ApJ...728..103L}, whose fan takes the shape of a separatrix
dome (Fig.\ref{fig_domes}a).

\begin{figure}[!h]
\centering \includegraphics[width=10cm]{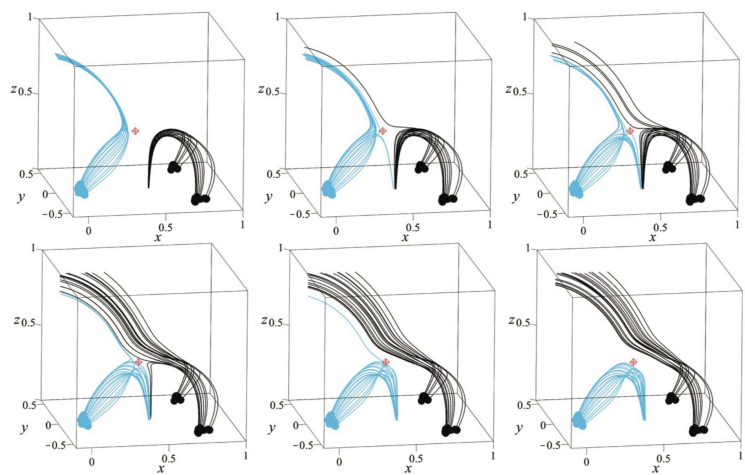}
\caption{Flux transfer for spine-fan reconnection at a null point (the red diamond), with
blue and black lines traced from fixed footpoints.  Blue field lines initially outside the separatrix surface
are transferred to inside the separatrix, while black field lines inside the separatrix  are moved
to outside the separatrix (from Pontin et al.\cite{Pontin2013ApJ...774..154P} with permission).} \label{fig_pontin}
\end{figure}

Spine-fan reconnection (Sec.2\ref{sec2.5}) has been  studied in MHD
simulations by Pontin et al.\cite{Pontin2013ApJ...774..154P}, who
demonstrated that magnetic collapse near a null  forms  a current
sheet localized around it. Flipping (or slipping or slip-running) of
field lines then occurs during the spine-fan reconnection, with the
magnetic connectivity of field lines continually changing
\cite{Aulanier2006SoPh..238..347A} and flux being transferred
between topologically different domains (Fig.\ref{fig_pontin}). The
flipping velocity becomes infinite on field lines that pass through
the null itself.

\subsubsection{Null Points and Solar Flares}
\label{sec3.1.1}
\begin{figure}[!h]
\centering \includegraphics[width=11cm]{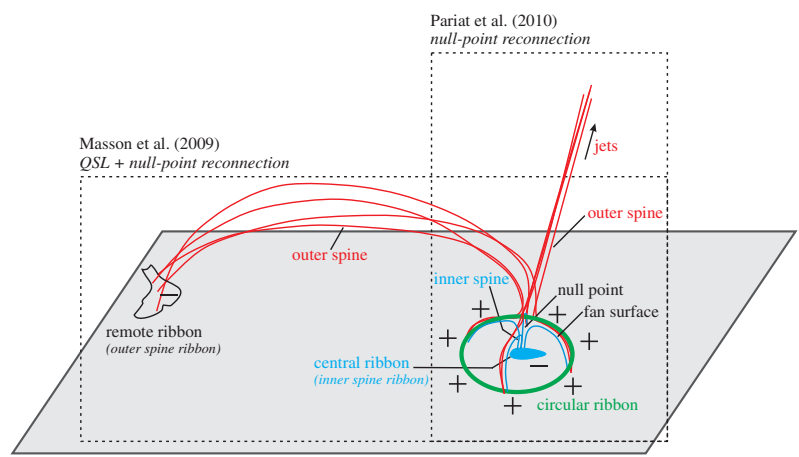}
\caption{Schematic of the relationship between circular flare ribbons, jets, and remote brightenings in a 3D fan-spine magnetic topology (from Wang \& Liu\cite{WanghM2012ApJ...760..101W} with permission).} \label{fig_wang}
\end{figure}

Null-point reconnection is thought to play a key role in many solar flares, especially circular-ribbon flares
\cite{Masson2009ApJ...700..559M,
Yangsh2020ApJ...898..101Y}.
Recent observations and 3D simulations show that when a magnetic bipole emerges into a unipolar region,
reconnection between the two fields forms a coronal null  configuration (with a 3D fan-spine structure) in
the corona\cite{
Liuw2011ApJ...728..103L,Houyj2019ApJ...871....4H
}. Null-point reconnection then generates heat and fast particles,
which travel along the fan separatrix and light up the fan
footpoints to form circular ribbons \cite{WanghM2012ApJ...760..101W}
(Fig.\ref{fig_wang}). In addition, a central ribbon and a remote
ribbon are observed at the footpoints of the spines that are located
below and above the dome, respectively.

Magnetic  extrapolations of the photospheric  field confirm that
twisted flux ropes are often present under the fan surface
\cite{
Songy2018ApJ...854...64S}.
When the flux rope loses equilibrium (due to nonequilibrium or kink
or torus
instabilities\cite{Hood1981GApFD..17..297H,Kliem2006PhRvL..96y5002K}),
the flux rope  rises  and triggers more violent null-point
reconnection\cite{SunX2013ApJ...778..139S
,ZhangQ2016ApJ...827...27Z}.
Such flux rope eruption can generate a blowout
jet\cite{Moore2010ApJ...720..757M
},
collimated from low down in the atmosphere
\cite{Raouafi2016SSRv..201....1R
}.

Circular flare ribbons often brighten sequentially in a clockwise or anti-clockwise direction \cite{Masson2009ApJ...700..559M,YangK2015ApJ...806..171Y
,LiT2018ApJ...859..122L
}.
For example, Li et al.\cite{LiT2018ApJ...859..122L} found circular ribbon elongation at a high speed of 220 km s$^{-1}$. It is a natural consequence of the flipping or slipping of magnetic field lines that occurs in null-point
reconnection, as demonstrated by Pontin et al \cite{Pontin2013ApJ...774..154P}.

\begin{figure}[!h]
\centering \includegraphics[height=1.5in]{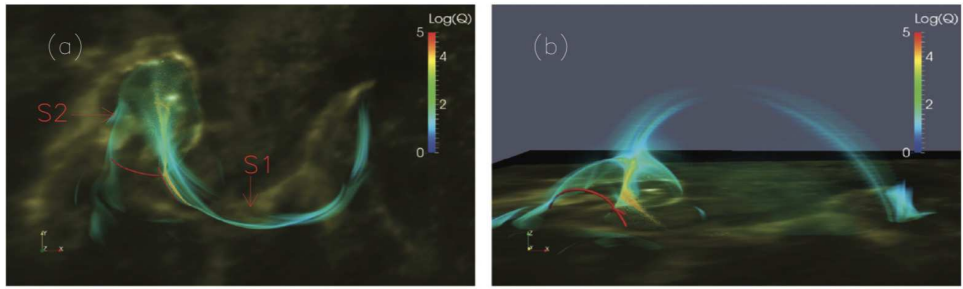}
\caption{A good agreement between the location of flare ribbons in 1600 {\AA} and values of Q for a potential field extrapolation.
Red lines denote the skeleton structure of the null point. S1 and S2 represent the western and eastern structures surrounding
the spine and fan field lines where Q is large (from Yang et al.\cite{YangK2015ApJ...806..171Y} with permission).} \label{fig_yang}
\end{figure}
Several authors have calculated the distribution of the squashing degree Q in configurations with coronal nulls or separators \cite{Masson2009ApJ...700..559M,YangK2015ApJ...806..171Y} (Fig.\ref{fig_yang}). Of course,
Q is infinite at the separatrices, although methods to determine Q will just show it to be high rather than infinite, due to the finite resolution of the methods. Nevertheless, when nulls or separators are present, they
give rise to null-point or separator reconnection rather than quasi-separator reconnection.
Pontin et al.\cite{Pontin2016SoPh..291.1739P} clarified this point by  showing that an extended high Q-halo around the spine or fan is a generic feature of null-point or separator reconnection. Thus, we stress that it is important to determine carefully whether there are any nulls or separators present before calculating Q,
since by itself neither Q nor the presence of flipping will distinguish between the different types of reconnection.  In addition, maps of Q will also show up structures away from
the separatrices where the mapping gradient is large but where current sheets do not form.

Null-point or separator reconnection is also involved in the {\it breakout mechanism}  proposed by Antiochos et
al.\cite{Antiochos1999ApJ...510..485A} as a possible explanation for the initiation of some CMEs and eruptive solar flares in a quadrupolar  configuration. In
the breakout model, reconnection between a low-lying sheared core flux and a large-scale overlying flux system
enables   the core flux to "break out". Later, the
model was extended to a fully 3D system, with
two polarity inversion lines, a separatrix dome and a 3D null point
at the intersection of the separatrix surface and the spine field
lines\cite{Lynch2008ApJ...683.1192L}. The
 mechanism has also been
used to explain small-scale solar jets
\cite{
Raouafi2016SSRv..201....1R,Wyper2017
}.

\subsubsection{Null Points in Earth's Magnetosphere}
\label{sec3.1.2}

\begin{figure}[!h]
\centering
\includegraphics[width=\columnwidth]{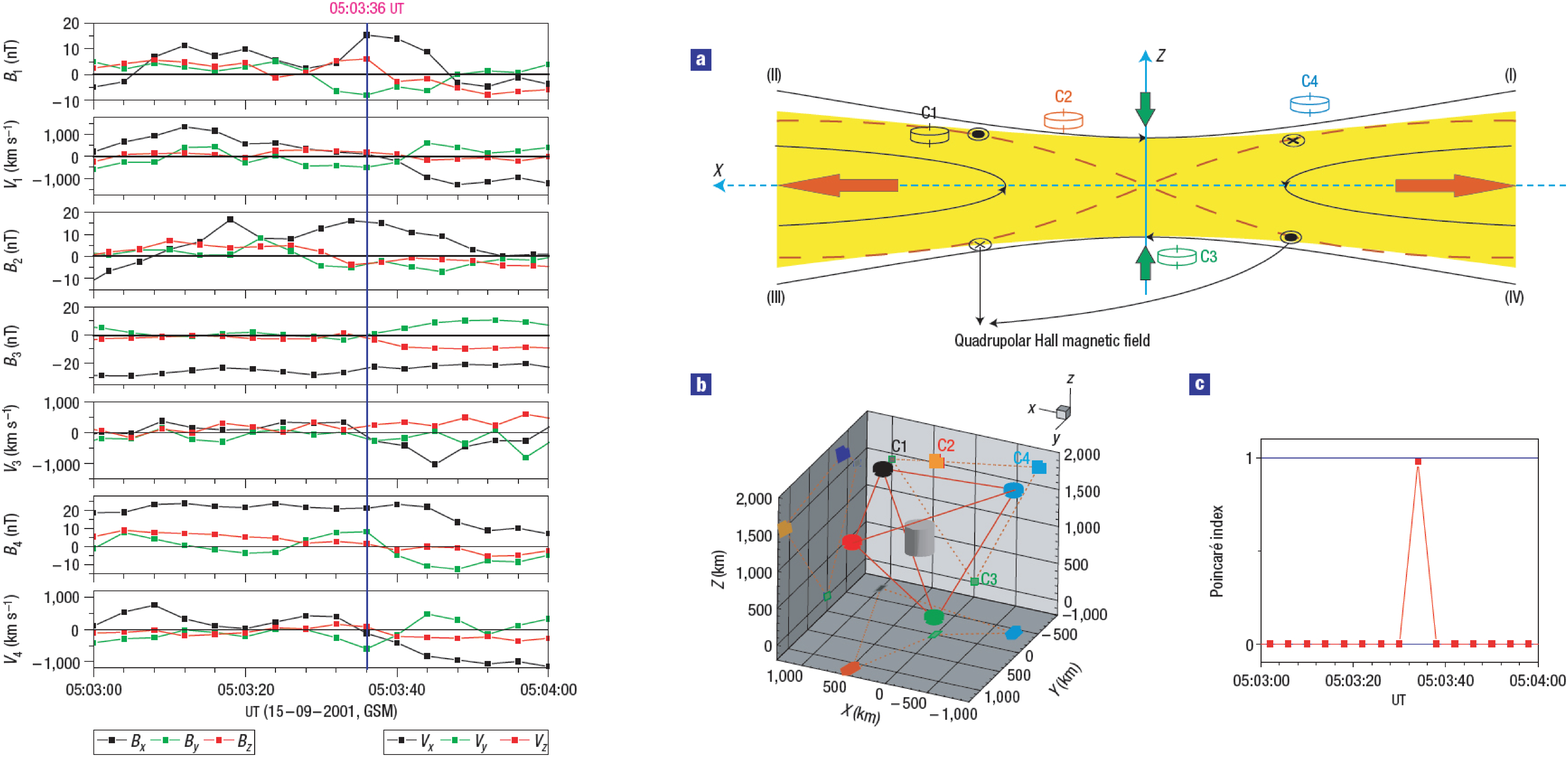}
\caption{{\it In situ} observations of a magnetic null point by four
Cluster spacecraft in a magnetotail reconnection region. (left) The
magnetic field and plasma flow. (right) The Poincar\'{e} index is
non-zero when the four spacecraft surround a null point. (From Xiao
et al. \cite{1216} with permission.)} \label{fig_firstobs}
\end{figure}
\begin{figure}[!h]
\centering\includegraphics[width=0.6\columnwidth]{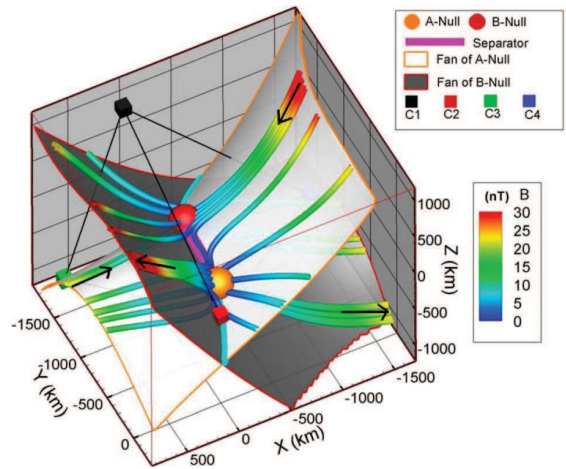}
\caption{A separator reconnection configuration with a magnetic null
pair reconstructed from Cluster measurements at 09:48:25.637 UT on
Oct 1st, 2001, consistent with the analysis of Xiao et al.
\cite{1217}. (From He et al. \cite{He08} with permission.)}
\label{fig_recons}
\end{figure}

{\it In situ} measurements of 3D magnetic null points have been made
in reconnection regions of the Earth's magnetosphere  (Fig.
\ref{fig_firstobs}) \cite{1216}, using observations from the four
Cluster spacecraft. They are identified from the Poincar\'{e} index
calculated from the observed magnetic vectors at four
non-coplanar points that surround the null point. The eigenvalues
and eigenvectors of the matrix $\delta B_{ij}=\nabla_jB_i$ are
calculated assuming a linear interpolation
and supposing there is only one null point in the tetrahedron
obtained from the four spacecraft positions. He et al.
\cite{he08b
} extended the  method to include an arbitrary number
of null points (Fig. \ref{fig_recons}). As a result,  3D null points
have been identified in the Earth's magnetotail \cite{1139,
 1217, 1216}, magnetopause \cite{809},  turbulent magnetosheath
\cite{
1119 }, and  bow shock \cite{2048}. The spatial scale for variations
of the magnetic field near the observed magnetic null
 is on the order of an ion
inertial length \cite{He08
}, implying that the Hall effect is
important.

Both spacecraft data and simulation results show that  spiral null points tend to occur more often than  radial null points
 \cite{2047,olshevsky16}, regardless of the
regions in the magnetosphere and the magnetosheath.
The observations at  ion scales show that spiral nulls are naturally
related to twisted magnetic flux ropes \cite{guo16, 1875, 1477},
which may play an important role in  plasma acceleration during 3D
reconnection \cite{1152
}. Indeed, energy dissipation is strong in outflow helical
field lines  \cite{2052} and in clusters of nulls \cite{868
}. The close relation between spiral nulls and flux ropes may well
hold also at other scales.

Cluster spacecraft data allow one to measure the plasma velocity,
plasma number density, electric field, and current flow near a null
point, as well as its motion. Perpendicular plasma flows around a
spiral null found by Wendel \& Adrian \cite{1119} often
rotate in the fan plane, especially when the fan or spine are
approached. The current is mainly along the spine but also has a
component perpendicular to the spine. The flows indicate a
combination of torsional spine and fan reconnection, which was also observed in the magnetotail  \cite{guo16}.

\begin{figure}[!h]
\centering\includegraphics[width=0.8\columnwidth]{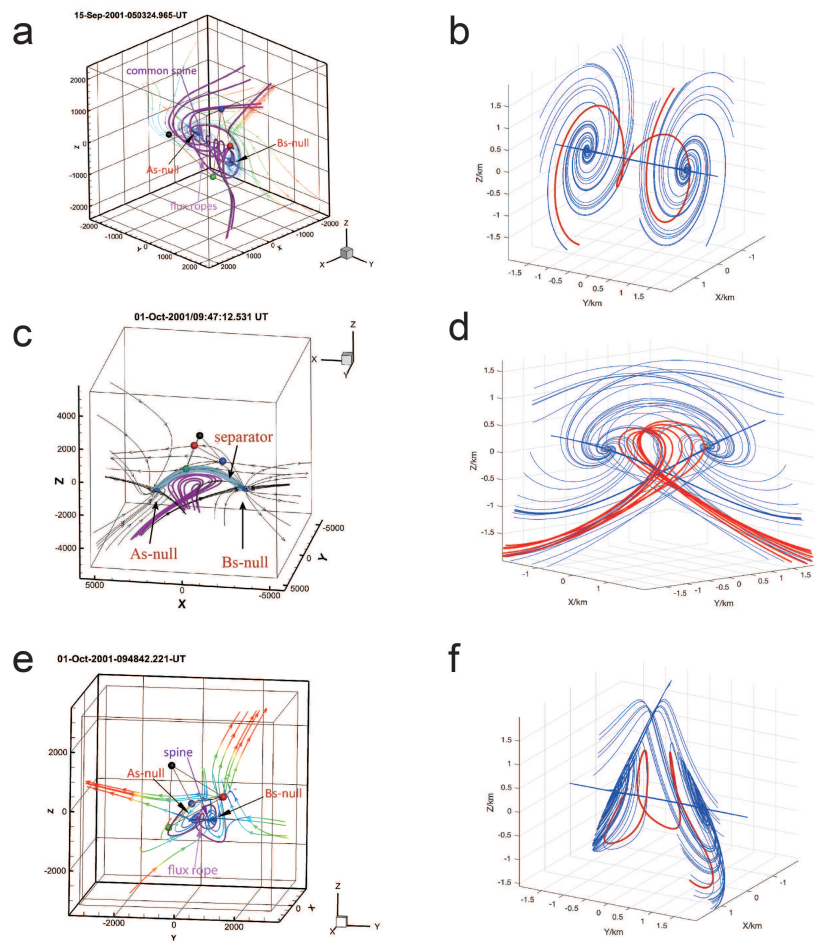}
\caption{(left) Reconstruction of three types of spiral null pair
observed by Cluster. (right) The configuration of the spiral null
pairs based on the analytical model with (b) $\varepsilon = 0$,
$\alpha = 0$, and $j = 6$; (d) $\varepsilon = 2$,  $\alpha = 3$, and
$j = 6$; (f) $\varepsilon = 0$,  $\alpha = 3$, and $j = 6$. (From
Guo et al. \cite{guo16} \cite{1875} with permission.)}
\label{fig_snullmodel}
\end{figure}
Nulls may be generated in pairs \cite{899} when a separator is born
(see Secs.2\ref{sec2.6},3\ref{sec3.2}). Spiral null pairs are
sometimes related to  the formation of flux ropes, when they are
chained by helical field lines along their spines. This structure
has been observed in the magnetotail by Cluster  \cite{guo16} and in
solar emerging flux simulations \cite{parnell10b}. Fig.
\ref{fig_snullmodel} gives the reconstructed magnetic structure for
three different spiral null pairs. In each case an As-type null is
joined to a Bs-type null (Sec.2\ref{sec2.2}). In Fig.
\ref{fig_snullmodel}a  the nulls
 are linked by their spine lines, which could be a common spine or more likely two helically wrapped spine lines, since the former is topologically unstable.
Field lines around the spines and between the two spiral nulls are twisted to form flux ropes. Such structures are also found in simulation results \cite{olshevsky16}, which show that the spiral
nulls are formed by the wrapping and kinking of a current sheet.

In Figs. \ref{fig_snullmodel}c and \ref{fig_snullmodel}e the nulls
are connected by a separator and by both spine and separator
\cite{1875}, respectively.
 These two structures, which were observed in the magnetotail, as well as the As-spine-Bs-like case can be represented by the following magnetic field \cite{1875}:
\begin{equation}
 (B_x,B_y,B_z\ )=[xy-\half\ jz+\varepsilon y,1-y^2+\alpha x,\ zy+\half \ jx].
\nonumber
\end{equation}
In this analytical model, the two spiral nulls are located along the $y$-axis and $j$ is the current density along the $y$-axis. The terms $\varepsilon y$ and $\alpha x$ represent  magnetic perturbations
parallel and perpendicular to the line joining the nulls.  The plots in the right column of Fig. \ref{fig_snullmodel} give three types of null pairs by choosing different values for $\varepsilon$ and $\alpha$. In
each structure, the flux ropes are formed between two spiral nulls and surround the spine lines. These different linkages between spiral null points could be generated by different bifurcation processes produced by
different magnetic perturbations  in time and space \cite{965}.

The formation, disappearance and bifurcation of  null points frequently takes place during 3D reconnection.
 As observed in the magnetotail, the number of  null points in the region enclosed by four Cluster
spacecraft can vary rapidly, presenting a turbulent-like
reconnection region \cite{868}. In a turbulent plasma, the
dissipation is largely produced by reconnection at clusters of null
points and short-lived radial null pairs and at the separators
joining the nulls \cite{parnell10b
}.

\subsection{Separator Reconnection: Modelling and Observations}
\label{sec3.2}
Separator reconnection is widely thought to be important in coronal heating, solar flares, and the Earth's magnetosphere, as described below (Secs. 3b\ref{sec3.2.2},3b\ref{sec3.2.3},3b\ref{sec3.2.4}). But
first we give examples of local and global skeletons of coronal magnetic fields constructed from photospheric magnetograms (Sec.3b\ref{sec3.2.1}).

After the crucial importance of analysing the {\it skeleton} of separatrix surfaces was realised \cite{priest96c
},  Parnell and Longcope  \cite{parnell08,parnell10b,haynes07,
longcope05b,longcope07a} developed the necessary techniques and applied them to numerical experiments and observed magnetic fields. The
importance of the skeleton and its evolution is that it reveals the different topological regions and how they evolve by separator reconnection, as well as how the separators are born, evolve, and disappear.
Later, Titov \cite{Titov2007ApJ...660..863T,titov09} generalised this concept to that of a {\it structural skeleton}, which is the sum of the skeleton and {\it quasi-skeleton} of quasi-separatrices (Sec.3\ref{sec3.3}).

\begin{figure}[!h]
\centering\includegraphics[width=0.5\columnwidth]{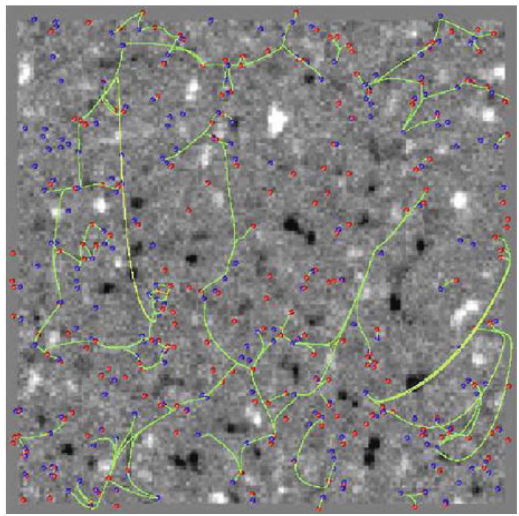}
\caption{Local skeleton from a potential field extrapolation above a photospheric magnetogram of a small part of the solar surface from SOHO/MDI, together with the positive nulls (red dots) and
negative (blue dots) nulls and separators (green curves). The observed photospheric magnetic field is of mixed polarity and highly fragmented and is known
as the {\it Magnetic Carpet}. (From Parnell et al.\cite{parnell11} with permission.)}
\label{fig_photmagm}
\end{figure}
\subsubsection{Skeletons from Photospheric Magnetograms}
\label{sec3.2.1}
The results from an early potential field extrapolation of a local photospheric magnetogram from SOHO/MDI are shown in Fig. \ref{fig_photmagm}, revealing the presence of many null points
produced by the highly localised and mixed-polarity nature of the magnetic flux protruding through the solar surface, known as the {\it magnetic carpet} \cite{schrijver98}.

\begin{figure}[!h]
\centering\includegraphics[width=0.9\columnwidth]{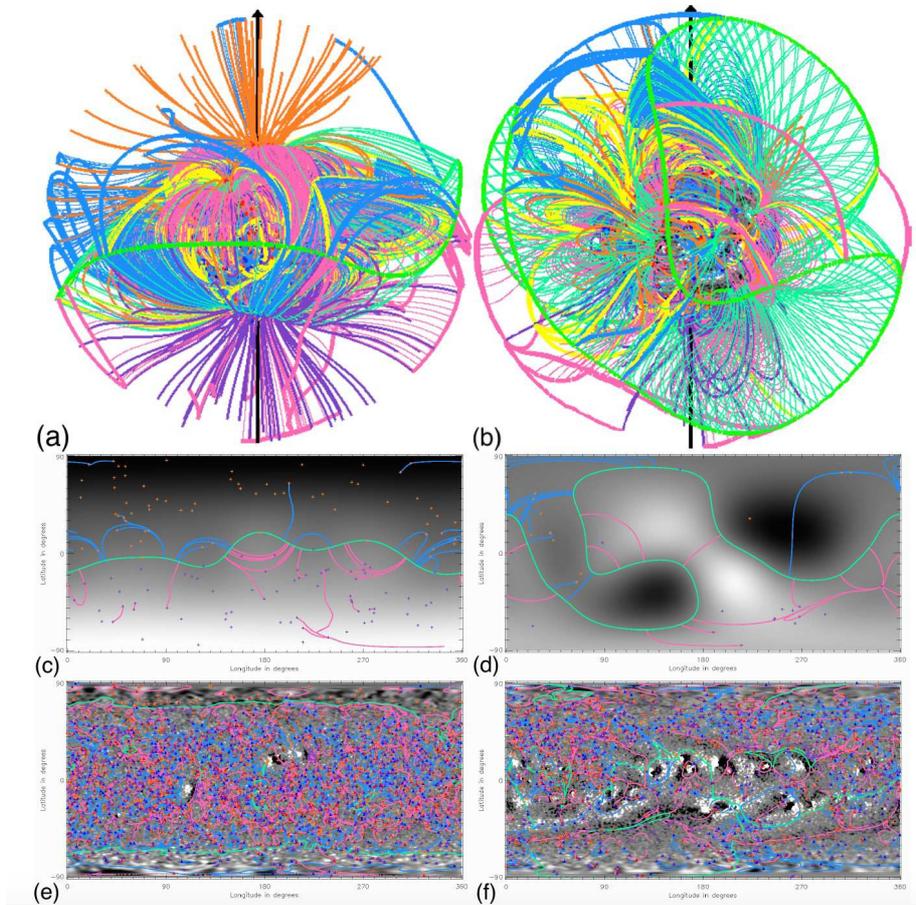}
\caption{Global skeleton from a potential field extrapolation above a synoptic photospheric magnetogram of the whole of the solar surface at (a) solar minimum and (b) solar maximum. The
topological features are the positive nulls (red dots) with spines (purple) and separatrices (thin pink lines), negative nulls (blue dots) with spines (orange) and separatrices (thin blue lines), spines,
and separators (green curves). Thick pink and blue lines denote where the separatrices meet the source surface ($r=2.5R_{\odot}$), whereas thick green lines mark the base of the heliospheric current
sheet and thin green lines extending down from them map out the heliospheric current sheet curtains dividing open and closed fields. (c) and (d) show  cuts at a radius $r=2.5R_{\odot}$, while (e)
and (f) show cuts at $r=1.005R_{\odot}$ as well as all the null points. (From Parnell et al.\cite{parnell15a} with permission.)}
\label{fig_globalskel}
\end{figure}
Later, the global coronal topology at solar minimum and solar maximum was calculated (Fig. \ref{fig_globalskel}) using SOLIS synoptic magnetograms and a global potential field model with a maximum
harmonic number of $l=301$ to extrapolate the magnetograms. The global study by Platten et al. \cite{platten14a} revealed 1964 nulls and 1946 separators at solar minimum, but 1131 nulls and 808 separators
at solar maximum. During solar minimum there are large areas of the photosphere with small-scale mixed polarity that create a highly complex network of nulls and separatrices (Fig. \ref{fig_globalskel}e).

Note that, for both local and global skeletons, much greater complexity with many more nulls and separators would be produced if much higher-resolution extrapolations from more recent magnetograms from SDO
and the SUNRISE balloon were undertaken.

\subsubsection{Separators and Coronal Heating}
\label{sec3.2.2}
Coronal heating due to several effects has been proposed. The idea that the corona is filled with myriads of current sheets that are continually forming and reconnecting to give {\it nanoflares} was proposed
by Parker \cite{parker72} and has traditionally been modelled in terms of braiding an initially uniform magnetic field  by footpoint motions \cite{galsgaard96,
pontin11a}.

However, the {\it flux tube tectonics model}  \cite{priest02a}
suggested that the magnetic carpet is crucial, since it highlights
the fact that the photospheric sources of coronal magnetic field are
not smoothly varying large-scale structures, but are instead highly
concentrated and localised magnetic fragments and intense flux tubes
(Figs.\ref{fig_photmagm},\ref{fig_globalskel}). The fact the
magnetic flux protrudes through the solar surface into the corona at
many small highly concentrated locations makes the chromospheric and
coronal magnetic field highly complex with myriads of null points
(or quasi-nulls) and separators (or quasi-separators) at which
current sheets can form and reconnection takes place
\cite{priest05a,demoortel06b}.

Thus, flux tube tectonics may be regarded as a modern development of Parker's nanoflare heating ideas which leads to much more efficient heating, since it considers the action not of complex photospheric
motions on a uniform field but of simple motions on a magnetic field that observations imply is highly complex.

A key way in which coronal tectonics heats the corona has been modelled in a pioneering numerical experiment by
Parnell, Galsgaard and Haynes \cite{galsgaard96,galsgaard00b, parnell04,haynes07,parnell10a}. They consider an elementary interaction between two photospheric magnetic
sources, in which one flux source moves past another flux source of opposite polarity in the presence of an overlying horizontal magnetic field that is so-called {\it flyby}. They found that, if the
separation of the sources is small enough, reconnection is driven at a series of separators.

\begin{figure}[!h]
\centering\includegraphics[width=0.8\columnwidth]{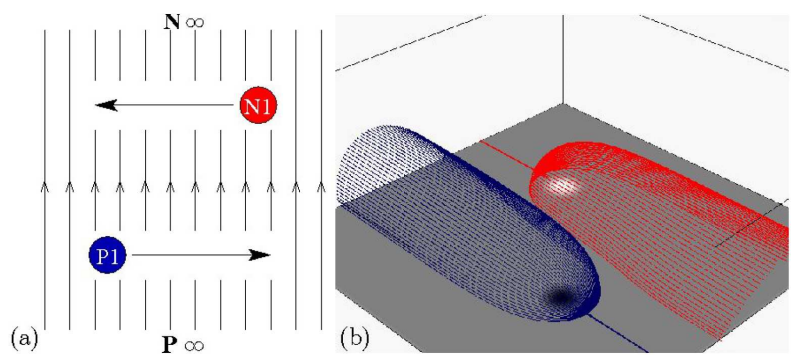}
\caption{Initial setup for a numerical flyby experiment showing (a) the view from above  and (b) the magnetic skeleton.  (From Parnell \& Haynes\cite{parnell09c} with permission.)}
\label{fig_flyby1}
\end{figure}
Fig. \ref{fig_flyby1}a shows the initial set up for Parnell's numerical experiment,  with two photospheric flux sources of opposite polarity ($P1$ moving to the right and $N1$ moving to the left) in an
overlying uniform field that is perpendicular to the motion of the sources. Figure \ref{fig_flyby1}b gives the initial magnetic skeleton, in which the two sources are not joined. There are two null points
in the photosphere, with fans that form two open separatrix surfaces extending along the direction of the overlying field. Below the blue separatrix all the flux from the positive source extends out through
one side boundary, while below the red surface the  flux from the negative source extends out through the opposite boundary. These fluxes are designated by the adjective {\it open}, whereas the flux that later
links one source to the other is called {\it closed}.

\begin{figure}[!h]
\centering\includegraphics[width=0.9\columnwidth]{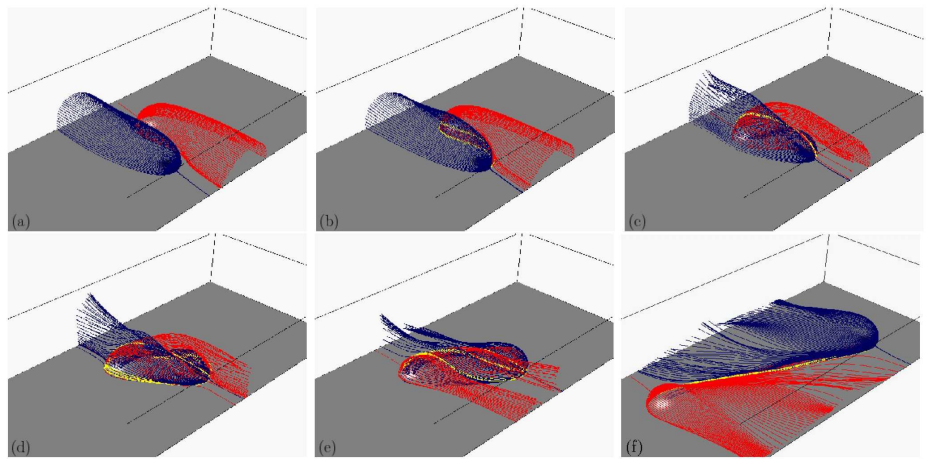}
\caption{The evolution of the skeleton for the flyby experiment, namely, the interaction of the two separatrix surfaces (red and blue) with separator field lines shown in yellow. (From Parnell \& Haynes\cite{parnell09c}
with permission.)}
\label{fig_flyby2}
\end{figure}
Then Fig. \ref{fig_flyby2} shows the subsequent evolution of the skeleton, in which the two separatrices intersect in a number of reconnecting separators that vary as the simulation proceeds.
\begin{figure}[!h]
\centering\includegraphics[width=0.9\columnwidth]{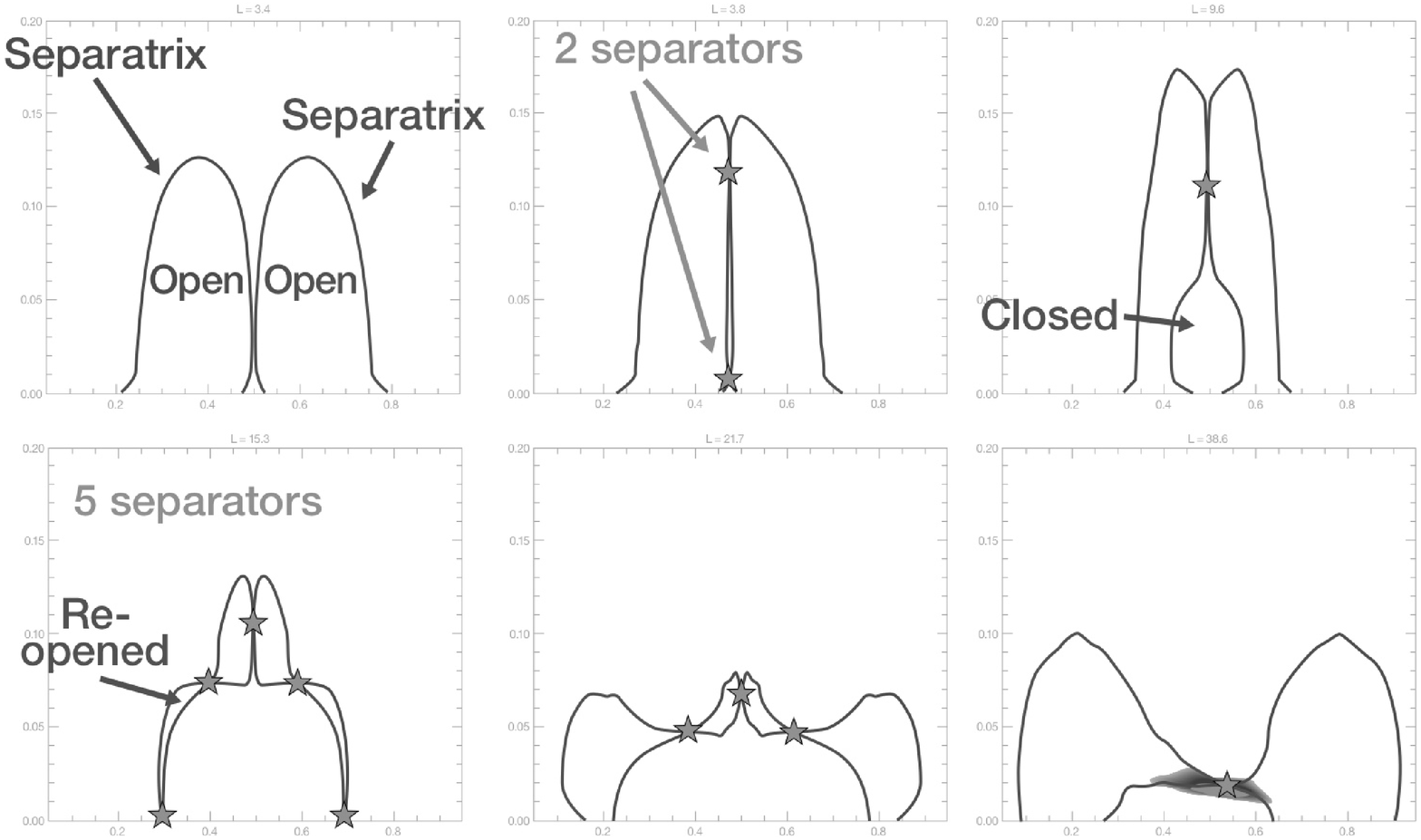}
\caption{The evolution of the intersections with a vertical cross-section at the mid-plane $y=0.5$ for the flyby simulation of Fig.\ref{fig_flyby2}. ``Open" refers to flux that links one of the
sources with a side boundary of the numerical box, while ``Closed" denotes flux that joins the two sources. (From Priest \cite{priest14a} with permission.)}
\label{fig_flyby3}
\end{figure}
It is. however, by taking a vertical section through the skeleton that what is happening becomes clear (Fig. \ref{fig_flyby3}). Initially, the two separatrices are completely separate, and then they touch
and intersect in two separators. One of these descends through the lower boundary to leave one separator, where reconnection builds up the closed flux linking the two sources. Next, the separatrix surface
that bounds the closed region touches the side separatrices and intersects them to give four more separators, two on each side. Two of these descend through the lower boundary to leave the central separator
and two side separators whose reconnection reopens part of the flux.

\begin{figure}[!h]
\centering\includegraphics[width=\columnwidth]{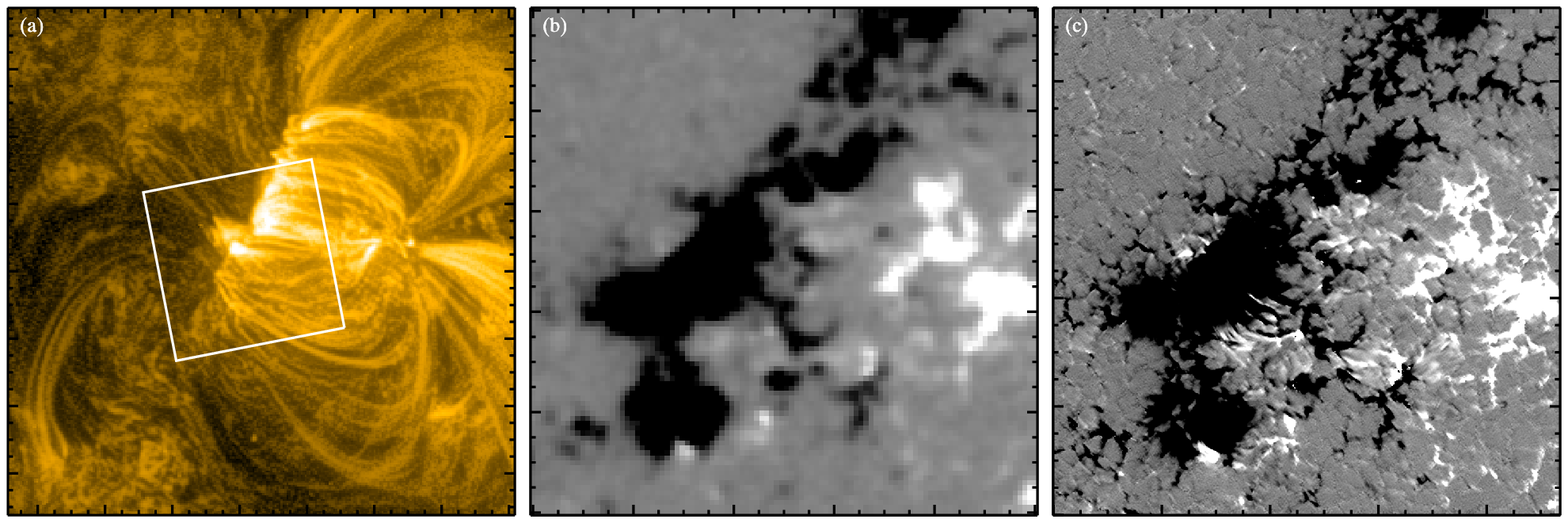}
\caption{ Coronal image of an active region on 2013 June 12 at 23:45
UT, and the underlying magnetic field. (a) An image from the
SDO/Atmospheric Imaging Assembly (AIA) 171 {\AA} filter in a 150x150
arcsec field of view. The white box covers an area of 51x51 arcsec
and encloses footpoint regions of several coronal loops. (b) SDO/HMI
magnetogram showing the distribution of the photospheric line of
sight magnetic field for the white box region of panel (a). (c) Same
as (b) but for the SUNRISE/IMaX observations. (From Priest et al.
\cite{priest18} with permission.)} \label{fig_sunrise}
\end{figure}
Recent observations from the SUNRISE balloon mission \cite{solanki10a} have revealed that the photospheric magnetic field is very much more complex than realised before and that magnetic
flux in the Quiet Sun is emerging and cancelling at a rate of 1100 Mx cm$^{-2}$ day$^{-1}$ \cite{smitha17}, which is an order of magnitude higher than thought before. Figs. \ref{fig_sunrise}a
and b show an image of the corona and the underlying magnetogram from SDO/HMI, which implies that the feet of the coronal loops are unipolar. However, Fig. \ref{fig_sunrise}c gives the equivalent much
higher-resolution magnetogram from SUNRISE, which reveals that the feet instead consist of many tiny regions of mixed magnetic polarity which are cancelling.

\begin{figure}[!h]
\centering\includegraphics[width=0.7\columnwidth]{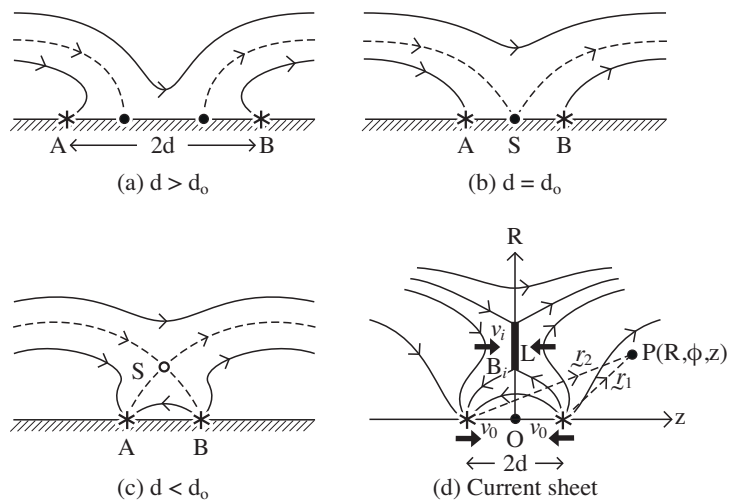}
\caption{The flux cancellation model for coronal heating. (a) Two
photospheric magnetic sources of flux $\pm F$, situated on the
z-axis a distance $2d$ apart in an overlying uniform horizontal
field $B_0$ approaching one another at speed $v_0$. (b) When $d =
d_0$, a separator S is formed. (c) Reconnection is driven at the
separator $S$ which rises in the atmosphere. (d) Energy is converted
at a current sheet of length $L$, where plasma flows in at speed
$v_i$ carrying magnetic field $B_i$. (From Priest \& Syntelis
\cite{priest21a}. Reproduced with permission from Astronomy \&
Astrophysics.)} \label{fig_cancellation}
\end{figure}
This led Priest et al. \cite{priest18} to propose a {\it flux cancellation model} for
chromospheric and coronal heating by nanoflares that are created not by braiding reconnection in the corona but by reconnection driven by photospheric flux cancellation (Fig. \ref{fig_cancellation}). As a
simple model of this process, they start with two opposite-polarity magnetic fragments of flux $\pm F$ in an overlying horizontal field $B_0$ and consider what happens as the half-distance $d$ between them
decreases and eventually the fragments cancel (Fig. \ref{fig_cancellation}). Initially, $d$ is large and there is no flux joining one fragment to the other, but, when $d$ becomes smaller than the interaction
distance \cite{longcope98}
\be
d_0=\left(\frac{F}{\pi B_0}\right)^{1/2},
\label{interacdist}
\ee
their fields start to reconnect at a separator. The length $L$ of the three-dimensional current sheet, as well as the inflow speed $v_i$ and magnetic field $B_i$, were calculated in terms of the
speed ($v_0$) of approach of the fragments, their fluxes ($F$) and the overlying field strength ($B_0$).

The energy release occurs in two phases: in the first phase, as the
fragments approach, the separator rises to a maximum height that
depends on $d_0$ and then falls to the photosphere; in the second
phase the fragments cancel. The maximum height of energy release can
be located in the chromosphere, transition region or corona,
depending on the parameter values, and in both cases energy is
released as heat and as the energy of a hot fast jet, as well as
fast particles. For observed parameter values, the energy release is
sufficient to heat the chromosphere and corona and can account for a
range of observed dynamic effects. More recently, the model has been
supported and extended by an analysis of reconnection at a 3D
separator current sheet \cite{priest21a}, as well as by numerical
simulations \cite{
syntelis20} and
observations of heating in the core of active regions in bright loops with flux cancellation at their footpoints \cite{chitta20}.

\begin{figure}[!h]
\centering \includegraphics[width=0.5\columnwidth]{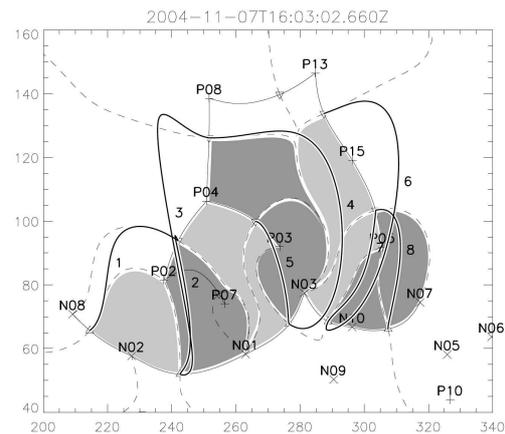}
\caption{The footprint of the skeleton of an active region,
indicating: sources that are positive (+) or negative ($\x$); null
points that are positive ($\del$) or negative ($\bm \Delta$); spines
(solid); footprints of fans (dashed); separators (thick curves); and
domains that gain flux (dark) or lose flux (light).
(From Longcope et al. \cite{longcope07e} with permission.)}
\label{fig_3flareskeleton}
\end{figure}
\subsubsection{Separators and Solar Flares}
\label{sec3.2.3} Separator reconnection is also a prime explanation
for many solar flares \cite{Longcope98a}. Longcope et al.
\cite{longcope07a,
Kazachenko12} have
shown how in many flares the stored energy can be  released by separator reconnection as it spreads through many domains of an active region, as described below.

Longcope et al. \cite{longcope07e}  predicted the flare energy release for several active regions and compared it favourably with observations. The coronal magnetic field is likely to  evolve through a series of nonlinear force-free equilibria with current sheets along separators, but these are difficult to calculate, and so Longcope \cite{longcope01} developed a simpler  {\it Minimum Current Corona (MCC)} model. In this model,
the photospheric magnetic field is split into a series of positive (P$_i$) and negative (N$_j$) unipolar flux patches, and the flux in the domain joining
P$_i$ to N$_j$ is calculated.
The domains are bounded by separatrices, which intersect in separators and make up the configuration's {\it skeleton} \cite{priest97b}.
In practice, reconnection between domains would conserve the total magnetic helicity and so the field in each domain would be force-free, but,
the MCC model assumes instead that the field evolves through a series of {\it flux-constrained equilibria} with the minimum energy that preserves the domain fluxes and with
 current sheets located along the separators. The free energy is then released by separator reconnection as flux is transferred between domains and the field is reduced to a potential one.

Longcope et al. \cite{longcope07e} partitioned a particular active region up into 28 regions, whose initial skeleton is shown in Fig.\ref{fig_3flareskeleton}a. It has 29 nulls and 32 separators.
They then calculated the changes in domain flux by a sequence of separator reconnections and so showed how reconnection  spreads through the region.
Flux changes are used to calculate the currents acting along each separator and also the energy released, which in this case amounts to $8\x 10^{31}$ erg, or 6$\%$ of the active-region energy. The  flare ribbons are found to lie along a series of spines that join a set of nulls.
Titov et al. \cite{titov12} applied the same idea to the triggering of a sequence of flares and CMEs.

\begin{figure}[!h]
\centering\includegraphics[width=0.6\columnwidth]{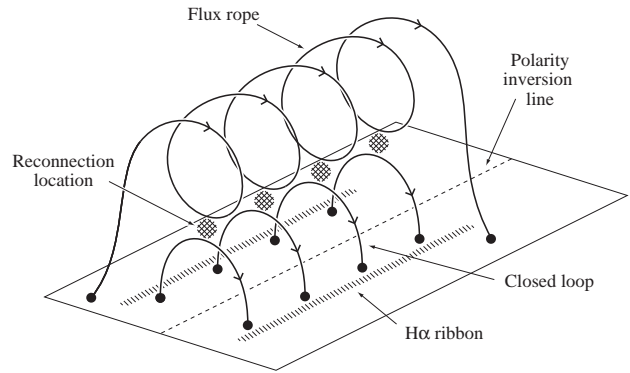}
\caption{A 3D flare cartoon for the creation of an arcade of flare loops and a twisted flux rope by reconnection at many sites above a polarity inversion line.
(From Longcope et al. \cite{longcope07e} with permission.)}
\label{fig_3Dflarecartoon}
\end{figure}
During a solar flare, 2D models suppose reconnection creates a closed field line or magnetic island, but 3D models instead imply that reconnection at a series of locations produces a twisted flux rope, as shown in
Fig.\ref{fig_3Dflarecartoon}, together with coronal arcade of flare loops.

\begin{figure}[!h]
\centering\includegraphics[width=0.6\columnwidth]{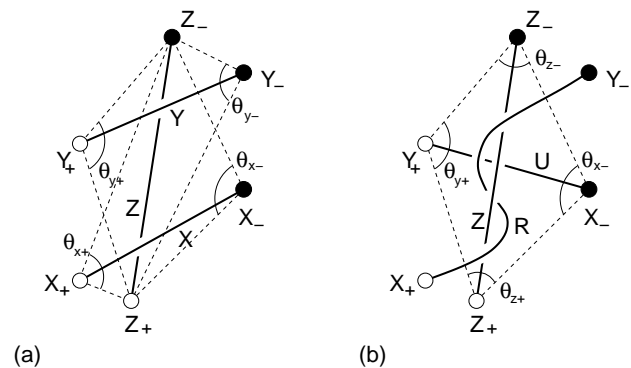}
\caption{The creation of twist by the zippette process, namely, reconnection of two coronal loops ($X_+X_-$,$Y_+Y_-$) overlying a prominence flux tube ($Z_+Z_-$) to create a twisted flux rope $X_+Y_-$ whose core is $Z_+Z_-$.
(From Priest \& Longcope \cite{priest17} with permission.)}
\label{fig_zippette}
\end{figure}
A model for {\it zipper reconnection} by Priest \& Longcope \cite{priest17} has addressed two important questions about the three-dimensional aspects of a flare, namely: during the rise phase,
how do two bright flare knots grow into ribbons while a single loop joining them develops into a flare arcade? and what is
the nature and magnitude of the resulting twist in the erupting flux rope? Fig. \ref{fig_zippette}a shows the {\it zippette process} in which two untwisted flux tubes (X$_+$X$_-$ and Y$_+$Y$_-$)
overlie an initial flux rope (Z$_+$Z$_-$), which reconnect below Z$_+$Z$_-$ to create an underlying flux tube U from Y$_+$ to X$_-$ together with a twisted flux rope R from X$_+$ to Y$_-$ that wraps around Z$_+$Z$_-$.

\begin{figure}[!h]
\centering\includegraphics[width=0.8\columnwidth]{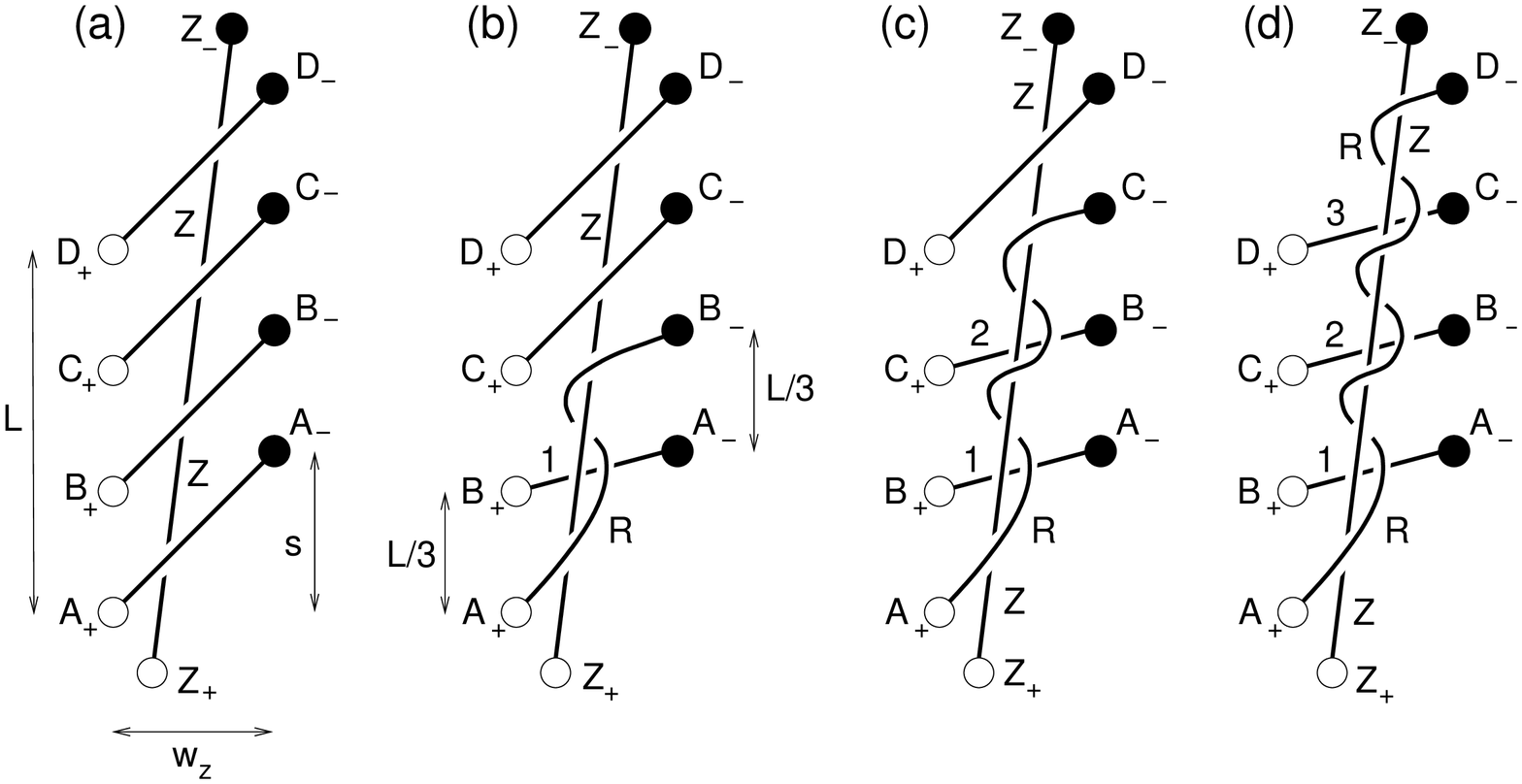}
\caption{A zipper model for the creation of flare ribbons and the build-up of twist in an erupting flux rope.
(From Priest \& Longcope \cite{priest17} with permission.)}
\label{fig_zipper}
\end{figure}
The idea in zipper reconnection (Fig.\ref{fig_zipper}) is then that, before the flare the magnetic configuration in an active region consists of an arcade of
coronal loops (A$_+$A$_-$, B$_+$B$_-$, C$_+$C$_-$, D$_+$D$_-$) overlying a filament or prominence Z$_+$Z$_-$,  whose magnetic field is a flux tube that may be untwisted or
only weakly twisted (Fig.\ref{fig_zipper}a). Then the flare is initiated when reconnection starts at one point in the arcade, such as, e.g., at the lower end in Fig.\ref{fig_zipper}a. Then, during the rise phase, zippette reconnection first takes place between A$_+$A$_-$ and B$_+$B$_-$ to produce a flux rope A$_+$B$_-$,
a flare loop B$_+$A$_-$ and brightening at the feet A$_+$B$_+$ and A$_-$B$_-$ as the start of the flare ribbons (Fig.\ref{fig_zipper}b). Next the reconnection spreads along the polarity inversion line,
gradually filling up the flare arcade and the flare ribbons. First of all, A$_+$B$_-$ reconnects with C$_+$C$_-$ to create the twisted rope A$_+$C$_-$ (Fig.\ref{fig_zipper}c),
and later A$_+$C$_-$ reconnects with D$_+$D$_-$ to create  A$_+$D$_-$.
At the end of this process (Fig.\ref{fig_zipper}d), the ribbons and arcade of hot loops have been created, together with a highly twisted flux rope,
whose core is the initial prominence field Z$_+$Z$_-$ and whose main part is A$_+$D$_-$. Once the flare ribbons have formed during the rise phase by zipper reconnection, the ribbons move apart during the main phase as reconnection moves to higher locations in the usual way.

Later, Priest \& Longcope \cite{priest20a} considered in detail the way twist is acquired by the 3D reconnection of two flux tubes and its distribution within the flux tubes.
One constraint on this process is that the total magnetic helicity is conserved as mutual helicity is converted to self-helicity and so creates twist. However, both a local and a global aspect to the process are also present:
the local effect is to produce equipartition of the amount of self-helicity and therefore twist that is added to the tubes; but the additional global effect implies that the location and orientation of the flux tube feet
generally add extra different self-helicities to the two tubes.

Two extra effects that are present in separator reconnection, but which have been highlighted in quasi-separator studies are the presence of hooks at the ends of the flare ribbons and the occurrence of flipping or
slipping of magnetic field lines. For more details, see Sec.3\ref{sec3.3}.

\subsubsection{Separators in the Earth's Magnetosphere}
\label{sec3.2.4} Separator reconnection is also critical in the
interaction between the solar wind and the Earth's magnetosphere.
This is a huge field and so we are reviewing only a small portion of
the literature here. Xiao et al. \cite{1217} showed for the first
time through {\it in situ} observations how the separator of a null
pair can serve as the location for reconnection. Also, the magnetic
configuration was determined and  a Hall electric field measured at
the separator \cite{He08}.

\begin{figure}[!h]
\centering\includegraphics[width=0.4\columnwidth]{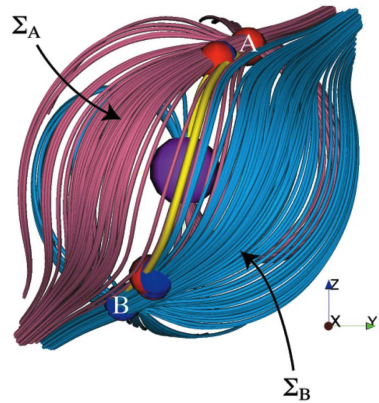}
\caption{Magnetic skeleton computed from the OpenGGCM simulation.
Type A (i.e., negative) nulls  are shown as red spheres, while Type
B (i.e., positive) nulls  are indicated by blue ones. The thick
yellow line lies approximately at the intersection of the two
separatrix surfaces ($\Sigma_A$ and $\Sigma_B$). (From Dorelli et al. \cite{1209} with
permission.)} \label{fig_mpsim}
\end{figure}
A long-lasting debate about magnetopause reconnection concerns the location of the line where reconnection takes place.
In 2.5D reconnection theory, two possibilities are antiparallel reconnection with no guide-field and component reconnection with a guide-field out of the 2D plane, but 2.5D is topologically unstable and so mention of antiparallel or component reconnection refers only to local behaviour without refence to the global 3D topology.
A 3D MHD simulation performed by Dorelli et al. \cite{1209} with zero magnetic dipole tilt and an IMF (interplanetary magnetic field) clock angle of 45$^{\circ}$ showed that null clusters form in two cusp regions connected by a separator that runs through the dayside magnetopause (Fig. \ref{fig_mpsim}), although such a separator is difficult to identify in {\it in situ} observations.
The simulation possesses two types of reconnection. The first is null-point reconnection in the cusp null clusters and in the high-latitude cusp region
(which a local 2.5D viewpoint would name antiparallel reconnection).
The second type is separator reconnection at the subsolar separator line (which a local 2.5D view would call
 component reconnection).

\begin{figure}[!h]
\centering\includegraphics[width=0.8\columnwidth]{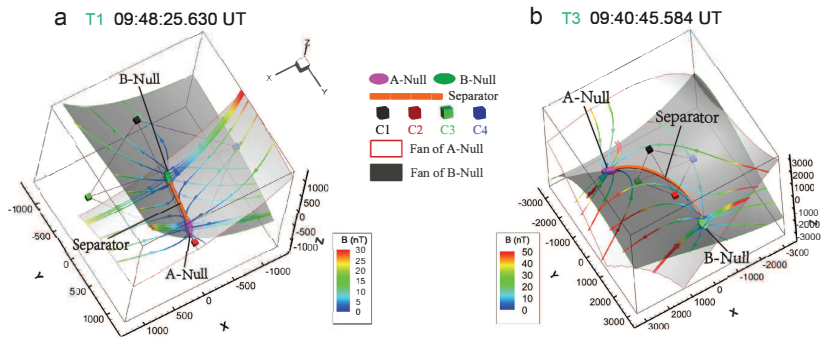}
\caption{(a) Reconstruction of (a) an antiparallel reconnection
region and (b) a component reconnection region.
(From Guo et al. \cite{160} with permission.)} \label{fig_apcp}
\end{figure}
Observations in the magnetotail also show that separator
reconnection can demonstrate features of both antiparallel and
component reconnection. Fig. \ref{fig_apcp} shows the reconstructed
magnetic structures for both cases \cite{160}. Radial null pairs and
the separator lines are present. For component reconnection, the
separator is long, and the four Cluster spacecraft cross the central
part of the separator line, where the magnetic strength is more than
10 nT by comparison with the ambient field strength of 20 -- 30 nT.
The large magnetic strength implies a large guide field. For
antiparallel reconnection, the two nulls are close to the spacecraft
tetrahedron and the maximum field strength along the separator is
only several nTs, which gives a small guide field.

\subsection{Quasi-Separator Reconnection: Modelling and Observations}
\label{sec3.3} 
\begin{figure}[!h]
\centering\includegraphics[width=0.8\columnwidth]{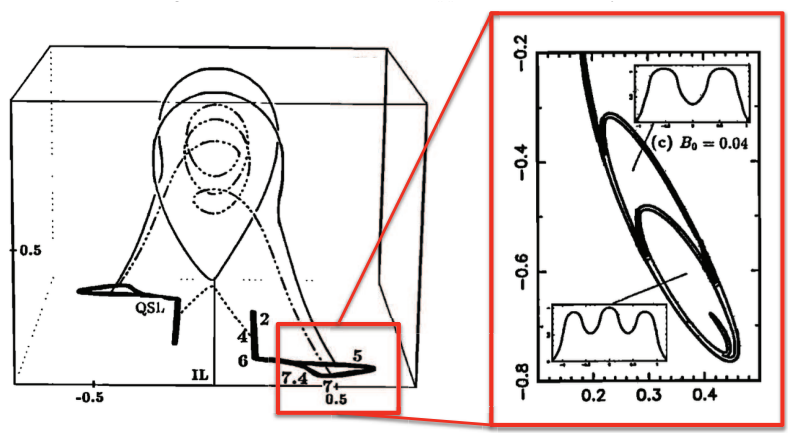}
\caption{QSL footprints associated with a flux rope and a zoom of
the hook-shaped part. Dash-dotted field lines denote the central
part of the twisted flux rope and
solid curves  the field lines at the periphery of the flux
rope. Dotted curves show the small arcade lying underneath the
flux rope. The QSL footprints form two elongated strips on both
sides of the inversion line. The footpoints of the twisted
flux rope are located inside the hook-shaped part of QSL footprints.
(From
D{\'e}moulin et al.\cite{Demoulin1996bJGR...101.7631D} with permission.)}
\label{fig_demoulin}
\end{figure}

Because of the strong distortion of the magnetic
field line mapping at QSLs, strong currents tend to build up at
quasi-separators, which, when no nulls or separators are present,
 are therefore preferential locations for
reconnection in general 3D systems\cite{priest1995,Demoulin1996}. Thus, in order to study
quasi-separator reconnection, it is important first to carefully
ensure that no nulls or separators are present. QSL locations are
found by measuring
 the squashing degree Q
\cite{Titov2002JGRA..107.1164T
}, which does not distinguish between separators and quasi-separators.
The role of quasi-separator reconnection in eruptive  and confined flares
has been widely investigated in numerical simulations
and observations.  In the analytical
work of D{\'e}moulin et al.\cite{Demoulin1996bJGR...101.7631D},
QSLs tend to wrap around flux ropes  and delineate the
frontier between different classes of field line (Fig.\ref{fig_demoulin}).
The intersection of QSLs with the lower boundary gives a shape that is typical of
observed flare ribbons\cite{Chandra2009SoPh..258...53C}.

After the notion of QSLs was proposed by Priest and D{\'e}moulin \cite{priest1995}, comparison of QSL
locations in flaring ARs with flare brightenings was carried
out by D{\'e}moulin and colleagues and others, using
a linear  force-free extrapolation \cite{Demoulin1997A&A...325..305D,
Mandrini2006SoPh..238..293M
}
or a nonlinear one \cite{Savcheva2012ApJ...750...15S,
ZhaoJ2016ApJ...823...62Z
}.
The photospheric traces of QSLs often
match well the locations of H$\alpha$ flare ribbons (often double J-shaped), while the 3D structure of QSLs and current concentrations resemble an S-shaped sigmoid and outline a flux rope.

\begin{figure}[!h]
\centering\includegraphics[width=0.32\columnwidth]{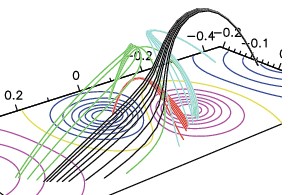}
\centering\includegraphics[width=0.32\columnwidth]{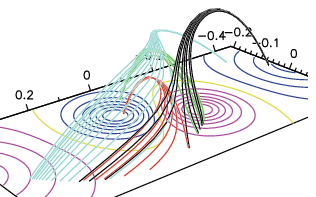}
\centering\includegraphics[width=0.32\columnwidth]{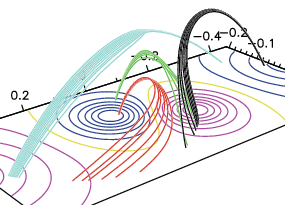}
\caption{Evolution of flipping or slipping field lines in a numerical
simulation. Pink
(blue) contours stand for positive (negative) polarity
magnetic fields. Four sets of magnetic field lines (red, black, cyan
and green lines) are integrated from fixed footpoints
and their conjugate footpoints gradually
flip or slip along  arc-shaped trajectories.
 (From Aulanier et
al.\cite{Aulanier2006SoPh..238..347A} with permission.)} \label{fig_aulanier}
\end{figure}
Flipping or slipping of field lines takes place in all 3D reconnection models. It
has been demonstrated clearly in 3D resistive
simulations on quasi-separator reconnection by Aulanier et
al.\cite{Aulanier2006SoPh..238..347A} in
 Fig.\ref{fig_aulanier}.
Reconnection naturally occurs along arc-shaped QSLs
and causes  red field lines to reconnect with black ones through
field line slippage. Ultimately, the initial red field lines
connecting the inner bipole and black field lines connecting the
outer bipole evolve into new lateral red and black field lines. A
similar process occurs between cyan and green field lines.

\begin{figure}[!h]
\centering\includegraphics[width=0.9\columnwidth]{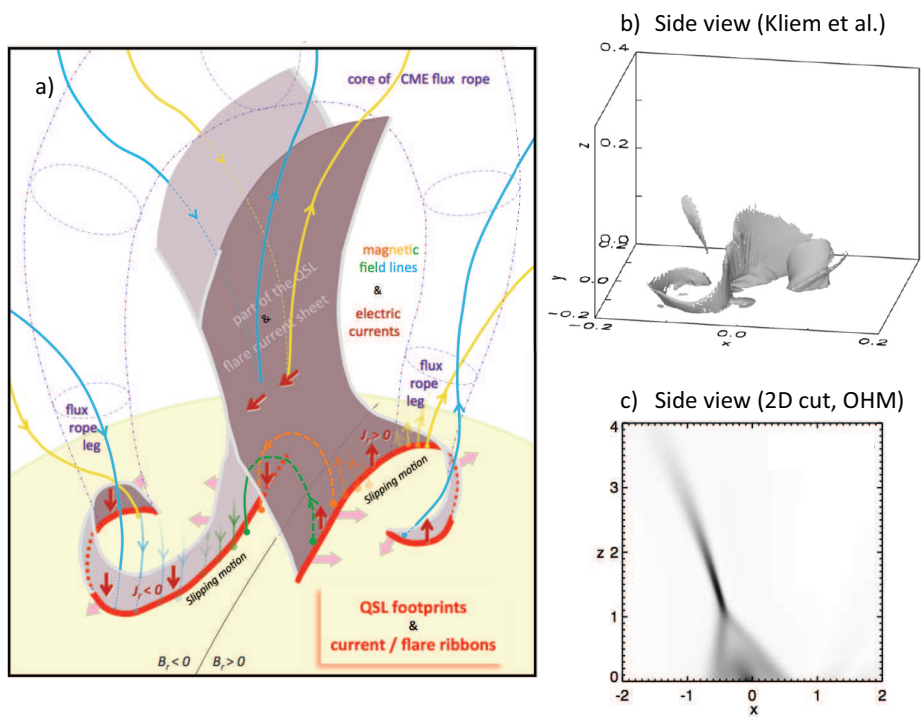}
\caption{ Panel (a) shows a A cartoon for a 3D standard model of
eruptive flares. The grey area indicates the QSL volume and a
current layer wrapping around a torus-unstable flux rope. The blue
and yellow loops form the outer envelope of the flux rope, and the
green and orange lines indicate the newly formed flare loops due to
the occurrence of reconnection. (From Janvier et
al.\cite{Janvier2014ApJ...788...60J} with permission.) Panel (b)
displays the isosurface of vertical current sheet that is formed
underlying the erupting flux rope. (From Kliem et
al.\cite{Kliem2013ApJ...779..129K} with permission.) Panel (c) shows
a 2D vertical cut of the electric current under the rising flux
rope, presenting a cusp shape. (From Janvier et
al.\cite{Janvier2013A&A...555A..77J}. Reproduced with permission
from Astronomy \& Astrophysics.)} \label{fig_janvier}
\end{figure}
Janvier et al.\cite{Janvier2013A&A...555A..77J} have simulated an
eruptive flare caused by a torus-unstable flux rope. They found a
linear correlation between the slippage speed and strength of the QSL. Based
on 3D MHD simulations  and
observations, Aulanier, Kliem and colleagues \cite{Aulanier2012A&A...543A.110A,
Janvier2014ApJ...788...60J
} proposed 3D extensions to the standard 2D CSHKP flare model
(Fig.\ref{fig_janvier}a). As the flux rope expands, regions of high
current density are formed along separatrices or QSLs (the grey
areas in Fig.\ref{fig_janvier}a), and the footprints of the current
layers give double J-shaped flare ribbons. Reconnection within the
current layers causes apparent slippage of flare ribbons, together
with a cusp-shaped region of high current density
(Fig.\ref{fig_janvier}c),  reminiscent of an HFT (Sec.2\ref{sec2.2})
\cite{Savcheva2012ApJ...750...15S,ZhaoJ2016ApJ...823...62Z}.

In the past 10 years, rich observations of flare loops and footpoints have become available with high
time and spatial resolution. Aulanier et
al.\cite{Aulanier2007Sci...318.1588A} reported
fast bidirectional slippage of coronal loops. SDO flare observations have revealed many features of   both 3D separator and quasi-separator reconnection. Dud{\'\i}k et
al.\cite{Dudik2014ApJ...784..144D} reported slipping
motions of both the flare and erupting loops along  developing
flare ribbons at speeds of tens of km
s$^{-1}$.

\begin{figure}[!h]
\centering\includegraphics[bb=27 268 539
545,clip,angle=0,scale=0.5,width=\columnwidth]{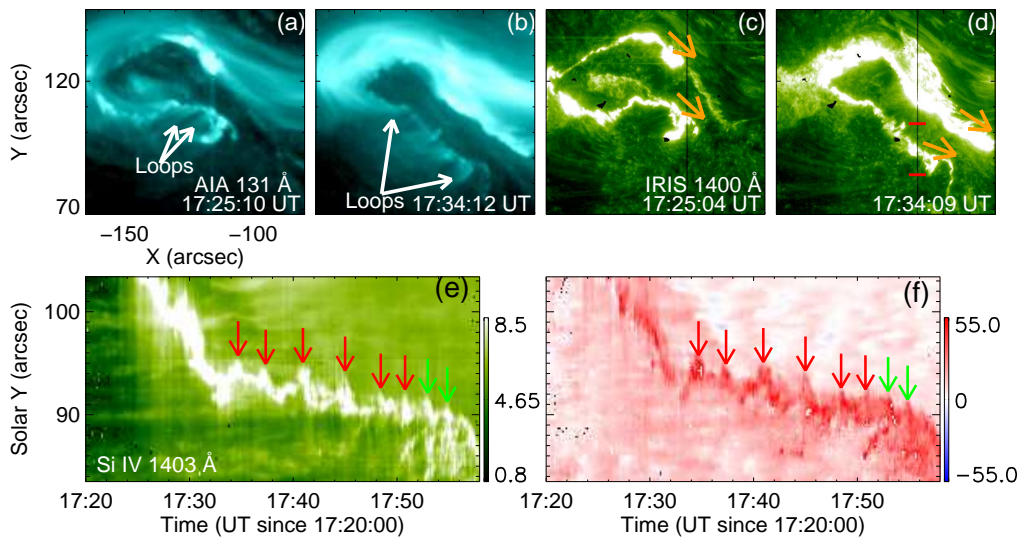}
\caption{Panels (a)-(d) show slipping hot flare loops and flare
ribbons as seen in SDO/AIA 131 {\AA} and IRIS 1400 {\AA}. The arrows
in panels (c)-(d) denote the slipping direction of flare ribbons.
Panels (e)-(f) display the temporal evolution of peak intensity and
Doppler shift in the spatial range indicated by the two red
horizontal bars in panel (d). These parameters are obtained by
applying single-Gaussian fits to the Si IV 1402.77 {\AA} line. The
arrows in panels (e)-(f) point to the peaks of wave-like evolution
and correspond to the times when the slipping knots pass by the
location of the IRIS slit.
 (From Li \&
Zhang\cite{LiT2015ApJ...804L...8L} with permission.)}
\label{fig_ting}
\end{figure}

Combining imaging and spectroscopic observations from SDO and IRIS,
Li \& Zhang\cite{LiT2015ApJ...804L...8L} found quasi-periodic
patterns with a period of 3-6 minutes in small-scale bright knots
that moved along  flare ribbons, while the flare loops exhibited
quasi-periodic slippage along the flare ribbon at a speed of 20-110
km s$^{-1}$ (Fig.\ref{fig_ting}).

Both separator and quasi-separator flare models suggest bi-directional slippage  along flare ribbons,
with one direction toward the ribbon hook building up the erupting flux rope and the
opposite direction  producing slippage of flare loops.
Dud{\'\i}k et al.\cite{DudIk2016ApJ...823...41D} observed flare loops slipping in opposite directions
at speeds of 20-40 km s$^{-1}$, but occasionally even faster velocities of 400-450 km s$^{-1}$ have been found
\cite{Zheng2016ApJ...823..136Z
}.
Also, Jing et al.\cite{Jing2017ApJ...842L..18J} observed a long-duration  flare ribbon slippage at a QSL footprint across a long distance ($\sim$60 Mm).

During  eruptive flares,  3D reconnection geometries are more complex
than in 2D. The photospheric footpoints, for example, are observed to brighten sequentially along the polarity inversion line during the rise
phase \cite{
Lit2014ApJ...791L..13L,
Chenh2019ApJ...887..118C,
Aulanier2019A&A...621A..72A}, as modelled by Priest and Longcope \cite{priest17}
in terms of zipper reconnection (Sec.3\ref{sec3.2}).
Also, Li et al.\cite{Lit2014ApJ...791L..13L} finds one end of  an eruptive flux rope is fixed and
the other end exhibits apparent slippage along a hook-shaped flare ribbon.
Aulanier \& Dud{\'\i}k\cite{Aulanier2019A&A...621A..72A} have proposed that a series of
reconnections between the flux-rope field lines and its surrounding arcade within
QSLs can produce a gradual drifting of  flux rope footpoints.

Moreover, both separator and quasi-separator reconnection have  been
shown to play a  role in confined solar flares\cite{Lit2018ApJ...869..172L,
Lit2020ApJ...900..128L}.
Li et al.\cite{Lit2018ApJ...869..172L} found  bi-directional slippage of ribbon
substructures along a ribbon in a confined flare. In fact, some confined flares are characterized by slippage and a stable filament,
whereas others possess a failed eruption of a filament or
flux rope\cite{
Lit2020ApJ...900..128L}. Just as for separators, quasi-separators may also play an important role in
coronal heating\cite{
Fletcher2001SoPh..203..255F,Schrijver2010ApJ...719.1083S}.
A study by Schrijver et al.\cite{Schrijver2010ApJ...719.1083S} compared
bright loops fanning out from ARs with topological features of QSLs, suggesting magnetic energy release
at separator or quasi-separator current sheets. Mandrini et al.\cite{Mandrini2015ApJ...809...73M}
found that persistent plasma upflows at the edge of ARs are often located near QSLs, and suggested
that reconnection  causes plasma to flow  from the high-pressure AR loops to neighboring large-scale
low-pressure loops in the quiet Sun.

\section{Conclusion}
\label{sec4}
Null points have been shown to be present in abundance in the solar
corona due to the complexity of  magnetic flux concentrated by
photospheric convective motions and projecting through the solar
surface into the atmosphere. The same is highly likely to be true in
other astrophysical environments such as the coronae of other stars
and of accretion disks, and nulls are also present in planetary
magnetospheres.

The fans of null points form a rich skeleton of separatrix surfaces
threading the corona, which intersect in separator field lines. The
dominant forms of magnetic reconnection, driven by photospheric
motions (in solar coronal heating) or magnetohydrodynamic
instability or nonequilibrium (in solar flares), are therefore
likely to be null reconnection or separator reconnection (and their
``quasi" equivalents).

During null or separator reconnection, the magnetic field lines
rapidly flip or slip through the plasma, while, at a null point or
separatrix, there is a discontinuity in the mapping of magnetic
field lines. Moreover, the behaviour is very similar at surfaces
called QSLs, where the mapping gradient of field lines is very large
but not singular, or in a region that one may call a quasi-null,
where the magnetic field becomes small but non-zero. Such QSLs
intersect in quasi-separators or HFTs.

Nulls, separatrices and separators are purely topological features, but, when the appropriate plasma flows are present, they are prime locations for the build-up of electric currents and therefore of
magnetic reconnection in diffusion regions.
Although quasi-nulls, QSLs and quasi-separators are not {\it topological} features, since there is no change of topology at them, they are important {\it geometrical} features of a complex magnetic
configuration. They often represent the remnants of nulls, separatrices and separators when a weak magnetic field is superposed to smooth away the null points.

However, just like their topological cousins, provided appropriate
plasma flows are present, currents will also build up near
quasi-nulls and quasi-separators, and so reconnection will take
place at them. Indeed, quasi-separator reconnection is also thought
to be important in solar coronal heating and solar flares, but
physically there is very little difference between it and separator
reconnection. The reason for the close similarity is simply that,
whereas reconnection takes place in strict 2D precisely at a null
point within a diffusion region, in 3D QSL reconnection or in 3D
null-point or separator reconnection the field lines continuously
change their connections everywhere throughout the diffusion region
that surrounds the null or separator.

Three-dimensional reconnection is a complex nonlinear process. However, its basic theory
(Sec.\ref{sec2}) has now been complemented by sophisticated numerical experiments that can go beyond the simplifying assumptions of theory and produce more realistic modelling of the
process (Sec.\ref{sec3}).  In addition, over recent years a new generation of solar telescopes has produced remarkable observations that have validated the basic theory and revealed a
wealth of detail on null, separator and quasi-separator reconnection in action.  These include the location of current concentrations and the sites of energy release, the presence of flipping or slipping of field lines,
and the creation of jets, both in coronal heating events and in solar flares.

For solar atmospheric heating, reconnection is likely to provide a major contribution, especially low down in the atmosphere and in active regions. In particular, the flux tube tectonics model is
a modern updating of the traditional nanoflare model, and a promising new development inspired by ultra-high resolution magnetograms is the flux cancellation model for both chromospheric and coronal heating.

For solar flares, we would like to propose the following new 3D
paradigm, which brings together a wide range of properties that have
been studied separately. The first two are taken over from the
standard 2D paradigm, while properties (iv,v,vii)  arise from
studies of separator reconnection \cite{longcope07a,
Kazachenko12,titov12,priest17,priest20a} and properties (viii,ix)
from quasi-separator reconnection
\cite{Lit2014ApJ...791L..13L,Janvier2014ApJ...788...60J,Mandrini1991},
but they apply equally to both separator and  quasi-separator
reconnection. \vspace{0.2cm}

 (i) A magnetic flux rope erupts due to magnetic nonequilibrium or instability and forms a vertical current sheet below the rope;

 (ii) Reconnection in the current sheet creates a rising arcade of flare loops with separating chromospheric ribbons at their feet as the height of the reconnection location increases;

 (iii) At low spatial and temporal resolution, reconnection may appear to
be quasi-steady and laminar,  especially during the late stages of a
flare, but, at high resolution, especially during the impulsive
phase, it is often impulsive and bursty in time and fragmented in
space, as revealed even during the main phase by observed
supra-arcade downflows \cite{mckenzie09,
  longcope18a};

 (iv) Reconnection starts at one location in the current sheet and creates two kernels of chromospheric emission;  then, during the rise phase, it spreads along the sheet above the
 polarity inversion line, gradually energizing the whole coronal arcade and forming the flare ribbons by zipper reconnection \cite{priest17}; the ribbons then move apart during the main phase;

 (v) Some of the twist in an erupting flux rope was present before the eruption, but most of it is created during the process of reconnection by the conversion of mutual magnetic helicity into self-helicity \cite{priest20a};

 (vi) The two main types of reconnection occurring in flares are null-point reconnection, which forms flare ribbons that are roughly circular, and separator (or quasi-separator) reconnection,
 whose ribbons are often straight or S-shaped;

 (vii) The topology (or quasi-topology) of an active region can be highly complex and partitioned into many domains bounded by separatrix (or quasi-separatrix) surfaces. Reconnection between different
 domains occurs at separators (or quasi-separators) and allows the energy release to spread from one domain to another, with the flare ribbons following a sequence of spines (or quasi-spines) \cite{longcope07a};
 in the same way, a series of coronal eruptions may be explained as a series of separator (or quasi-separator) reconnections between one region and another \cite{titov12};

 (viii) Flare ribbons often have hook-like ends \cite{Lit2014ApJ...791L..13L}, which are the ends of flux ropes bounded by quasi-separatrix (or separatrix) surfaces \cite{Janvier2014ApJ...788...60J};

 (ix) Flipping or slipping of magnetic fields, which was predicted to be a property of all 3D reconnection models, is observed in the behaviour of flare loops and their footpoints \cite{Mandrini1991}.
\vspace{0.2cm}

In the Earth's magnetosphere, observations with four spacecrafts can
better identify the presence of
 null points and separators and can directly obtain magnetic and plasma properties around them.
Spiral null points occur more frequently
 than radial null points. The former can
serve as the skeleton of magnetic flux ropes and play an important
role in magnetic energy release and plasma acceleration in the 3D
reconnection region. At the reconnection site,  separator
reconnection can demonstrate  features of both antiparallel
reconnection and component reconnection. Additionally, the formation
and evolution of clusters of null points are related to magnetic
turbulence  and significant energy dissipation.

It is clear that in future, developments in the basic theory,
numerical simulations and high-resolution observations will continue
to complement one another and to spur each other on to a fuller
understanding. It is also clear that the main distinction in types
of reconnection is between null-point reconnection and separator
reconnection, and that the distinction between separator and
quasi-separator reconnection is a largely theoretical distinction
that is unimportant for the physical consequences and observations
of energy release. Furthermore, likely future developments concern
unsteady patchy reconnection and better links between a macroscopic
MHD understanding and a microscopic collisionless understanding.
\enlargethispage{20pt}

\ethics{
There are no ethics issues.}
\dataccess{This article has no additional data.}

\aucontribute{All the authors have contributed equally to all
sections.}

\competing{There are no competing interests.}

\funding{This research is supported by the Strategic Priority
Research Program of Chinese Academy of Sciences (XDB41000000), the
National Natural Science Foundations of China (11773039, 11903050,
11790304, 11790300 and 41704169), the National Key R\&D Program of
China (2019YFA0405000), Key Programs of the Chinese Academy of
Sciences (QYZDJ-SSW-SLH050), the Youth Innovation Promotion
Association of CAS (2017078) and NAOC Nebula Talents Program. R.L.G.
is supported by the Incoming Post-Docs in Sciences, Technology,
Engineering, Materials and Agrobiotechnology (IPD-STEMA) project
from Universit\'{e} de Li\`{e}ge.}

\ack{E.P. is grateful to Guillaume Aulanier, Pascal D{\'e}moulin,
Terry Forbes, Gunnar Hornig, Dana Longcope, Clare Parnell, David
Pontin and Slava Titov for teaching him so much about the intriguing
nature of reconnection.}

\disclaimer{The ideas presented here represent our own ideas together with our understanding of those of other researchers.}




%
%
%
%
%





\bibliographystyle{RS}
\bibliography{sample}

\end{document}